\documentclass[11pt,onecolumn,aip,apl,reprint]{revtex4-1}

\renewcommand{\thesubsection}{\arabic{subsection}}

\newcommand*{\citen}[1]{%
	\begingroup
	\romannumeral-`\x 
	\setcitestyle{numbers}%
	\cite{#1}%
	\endgroup   
}

\newcommand{\addDS}[1]{\textcolor{black}{#1}}

\usepackage{graphicx,amsmath,amssymb,color}
\usepackage[outdir=./]{epstopdf}
\usepackage{afterpage}
\usepackage{textcomp}
\usepackage{xr}
\usepackage{float}
\usepackage{booktabs}

\usepackage{titlesec}
\titlespacing*{\section}
{0pt}{1.5ex plus 1ex minus .2ex}{1.5ex plus .2ex}
\titlespacing*{\subsection}
{0pt}{1.5ex plus 1ex minus .2ex}{1.5ex plus .2ex}

\newcommand{\sous}[1]{\ensuremath{_{\textrm{#1}}}}
\begin{document}
\title{Resonant terahertz detection using graphene plasmons}

\author{Denis A. Bandurin}
\email{bandurin.d@gmail.com}
\affiliation{School of Physics, University of Manchester, Oxford Road, Manchester M13 9PL, United Kingdom}

\author{Dmitry Svintsov}
\affiliation{Moscow Institute of Physics and Technology (State University), Dolgoprudny 141700, Russia}

\author{Igor Gayduchenko}
\affiliation{Physics Department, Moscow State University of Education (MSPU), Moscow, 119435, Russian Federation}
\affiliation{Moscow Institute of Physics and Technology (State University), Dolgoprudny 141700, Russia}

\author{Shuigang G.~Xu}
\affiliation{School of Physics, University of Manchester, Oxford Road, Manchester M13 9PL, United Kingdom}
\affiliation{National Graphene Institute, University of Manchester, Manchester M13 9PL, United Kingdom}

\author{Alessandro Principi}
\affiliation{School of Physics, University of Manchester, Oxford Road, Manchester M13 9PL, United Kingdom}

\author{Maxim Moskotin}
\affiliation{Physics Department, Moscow State University of Education (MSPU), Moscow, 119435, Russian Federation}
\affiliation{Moscow Institute of Physics and Technology (State University), Dolgoprudny 141700, Russia}

\author{Ivan Tretyakov}
\affiliation{Physics Department, Moscow State University of Education (MSPU), Moscow, 119435, Russian Federation}

\author{Denis Yagodkin}
\affiliation{Moscow Institute of Physics and Technology (State University), Dolgoprudny 141700, Russia}
\affiliation{Physics Department, Moscow State University of Education (MSPU), Moscow, 119435, Russian Federation}

\author{Sergey Zhukov}
\affiliation{Moscow Institute of Physics and Technology (State University), Dolgoprudny 141700, Russia}

\author{Takashi Taniguchi}
\affiliation{National Institute for Materials Science, 1‐1 Namiki, Tsukuba, 305‐0044 Japan}

\author{Kenji Watanabe}
\affiliation{National Institute for Materials Science, 1‐1 Namiki, Tsukuba, 305‐0044 Japan}

\author{Irina~V.~Grigorieva}
\affiliation{School of Physics, University of Manchester, Oxford Road, Manchester M13 9PL, United Kingdom}

\author{Marco Polini}
\affiliation{Istituto Italiano di Tecnologia, Graphene Labs, Via Morego 30, 16163 Genova, Italy}
\affiliation{School of Physics, University of Manchester, Oxford Road, Manchester M13 9PL, United Kingdom}

\author{Gregory N. Goltsman}
\affiliation{Physics Department, Moscow State University of Education (MSPU), Moscow, 119435, Russian Federation}

\author{Andre K. Geim}
\affiliation{School of Physics, University of Manchester, Oxford Road, Manchester M13 9PL, United Kingdom}
\affiliation{National Graphene Institute, University of Manchester, Manchester M13 9PL, United Kingdom}

\author{Georgy Fedorov}
\email{fedorov.ge@mipt.ru\\These authors contributed equally: D.A. Bandurin and D. Svintsov}
\affiliation{Moscow Institute of Physics and Technology (State University), Dolgoprudny 141700, Russia}
\affiliation{Physics Department, Moscow State University of Education (MSPU), Moscow, 119435, Russian Federation}

\begin{abstract}
	Plasmons, collective oscillations of electron systems, can efficiently couple light and electric current, and thus can be used to create sub-wavelength photodetectors, radiation mixers, and on-chip spectrometers. Despite considerable effort, it has proven challenging to implement plasmonic devices operating at terahertz frequencies. The material capable to meet this challenge is graphene as it supports long-lived electrically-tunable plasmons. Here we demonstrate plasmon-assisted resonant detection of terahertz radiation by antenna-coupled graphene transistors that act as both plasmonic Fabry-Perot cavities and rectifying elements. By varying the plasmon velocity using gate voltage, we tune our detectors between multiple resonant modes and exploit this functionality to measure plasmon wavelength and lifetime in bilayer graphene as well as to probe collective modes in its moir\'e minibands. Our devices offer a convenient tool for further plasmonic research that is often exceedingly difficult under non-ambient conditions (e.g. cryogenic temperatures) and promise a viable route for various photonic applications.
\end{abstract}

\maketitle

\begin{figure*}[ht]
	\centering\includegraphics[width=0.6\linewidth]{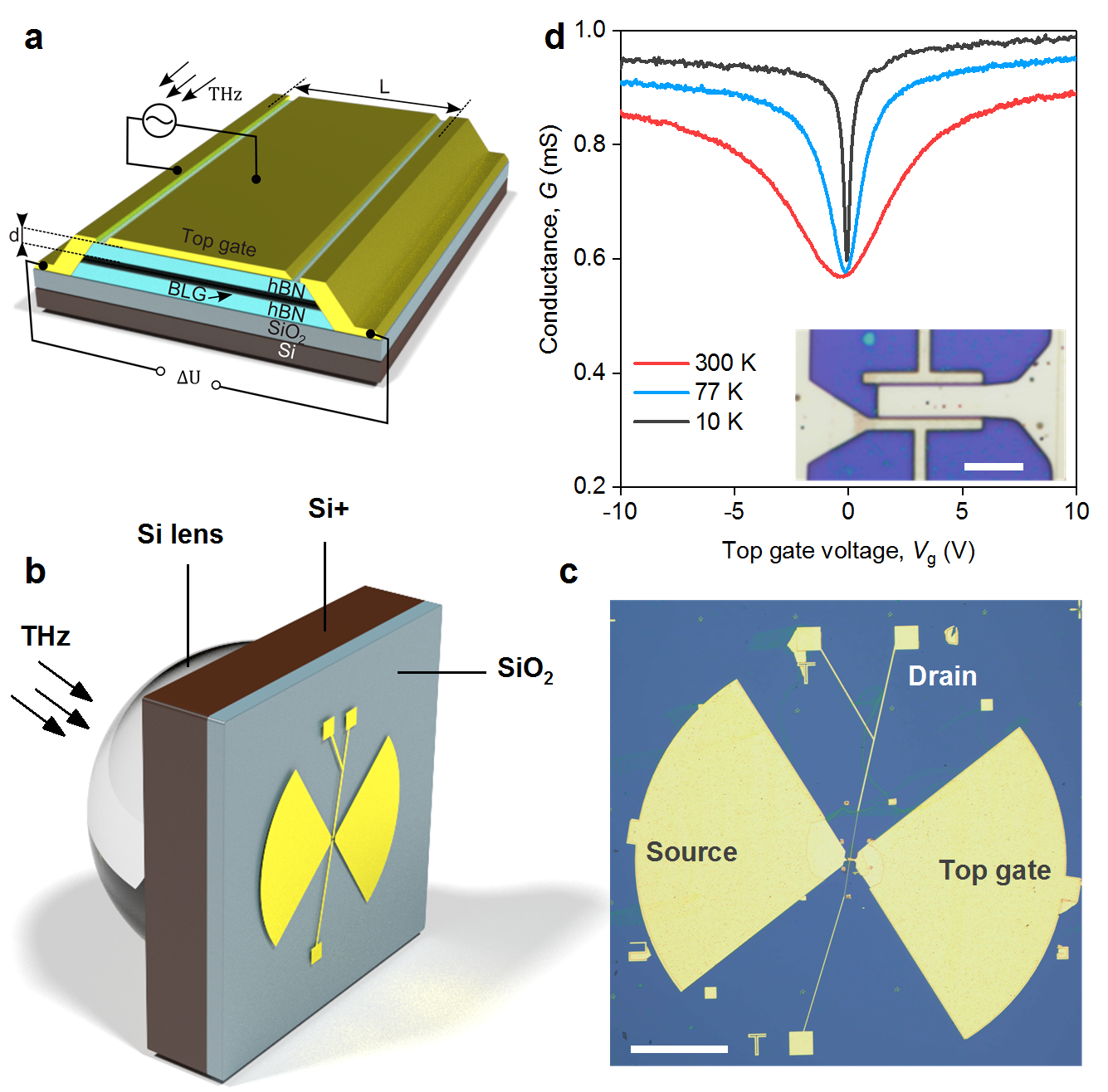}
	\caption{\textbf{Graphene-based THz detectors.} \textbf{a,}~Schematics of the encapsulated BLG FET used in this work. \textbf{b,}~3D rendering of our resonant photodetector. THz radiation is focused to a broadband bow-tie antenna by a hemispherical silicon lens yielding modulation of the gate-to-source voltage, as indicated in (a). \textbf{c,}~Optical photograph of one of our photodetectors. Scale bar is 200 $\mu$m.  \textbf{d,}~ Conductance of one of our BLG FETs as a function of the gate voltage $V\sous{g}$, measured at a few selected temperatures. Inset: Zoomed-in photograph of (c) showing a two-terminal FET with gate and source terminals connected to the antenna. Scale bar is 10 $\mu$m. }
	\label{Fig_1}
\end{figure*}		

Selective detection and spectroscopy of THz fields is a challenging task in modern optoelectronics offering a wide range of applications: from security and medical inspection to radio astronomy and wireless communications~\cite{Dhillon2017,Daryoosh2013}. Among the variety of available detection principles~\cite{Daryoosh2013}, one elegant proposal has always stood out and remained intriguing for more than two decades. The idea is to compress incident radiation into highly-confined two-dimensional plasmons propagating in the field effect transistor (FET) channel and to rectify the induced ac potential using the same device~\cite{Dyakonov1996a}. The FET channel, in this case, acts as a tunable plasmonic cavity with a set of resonant frequencies defined by its length and the density of charge carriers. The implementation of such resonant devices has promised on-chip selective sensing, spectroscopy, mixing, and modulation of THz fields below the classical diffraction limit~\cite{Dyakonov1996a}. However, despite decades-long experimental efforts, the excitation of long-lived plasma oscillations in conventional FETs has proven challenging~\cite{Knap2009, Vicarelli2012, Spirito2014,Tong2015, Qin2017, Generalov2017a} and little evidence of resonant THz detection has been found so far~\cite{Peralta2002,Knap2002,Teppe2005,Muravev2012,Giliberti2015}.

Graphene has recently demonstrated great promise for mid- and far-infrared plasmonics~\cite{Ju2011,Yan2012a,Fei2012,Chen2012,A.N.GrigorenkoM.Polini2012a,Woessner2015,Alonso-Gonzalez2016,Basov2018} and attracted a great deal of attention as a platform for plasmonic radiation detectors~\cite{LKoppens2014,A.N.GrigorenkoM.Polini2012a}. 
With lowering the operation frequency down to the THz domain, the resonant excitation of plasmons becomes exceedingly difficult and can only be achieved if the momentum relaxation rate is below the plasmon frequency, which, in turn, requires ultra-high electron mobility.
For this reason, in 
all graphene-based far-field THz detectors reported so far, the 
plasma waves -- if any-- were overdamped, and the devices exhibited only a broadband (non-resonant) photoresponse~\cite{Vicarelli2012,Cai2014, Spirito2014,Tong2015,Qin2017,Generalov2017a,Auton2017}. As a result, numerous applications relying on resonant plasmon excitation (see e.g. Refs.~\citen{Dyakonov1996a,Otsuji2006, Tomadin2013c,Ryzhii2012_resonantdetector,Fateev2017_rectification_periodic}) remain experimentally yet unrealized.

In this work, we demonstrate this long-sought resonant regime using FETs based on high-quality van der Waals heterostructures. In particular, we employ graphene encapsulated between hexagonal boron nitride (hBN) crystals which have been shown to provide the cleanest environment for long-lived graphene plasmons~\cite{Woessner2015,Basov2018}. Antenna-mediated coupling of such FETs to free-space radiation results in the emergence of dc photovoltage that peaks when the channel hosts an odd number of plasmon quarter-wavelengths. Exploiting the gate-tunability of plasmon velocity, we switch our detectors between more than ten resonant modes, and use 
this functionality to measure plasmon wavelength and lifetime. Thanks to the far-field radiation coupling, our compact devices offer a convenient tool for studies of plasmons in 
two-dimensional electron systems under non-ambient conditions (e.g. cryogenic environment and high magnetic fields) where other techniques may be arduous. As an example, we apply our approach to probe plasmons in graphene/hBN superlattices and unveil collective modes of charge carriers in moir\'e minibands.

\begin{figure*}[ht!]
	\centering\includegraphics[width=0.9\linewidth]{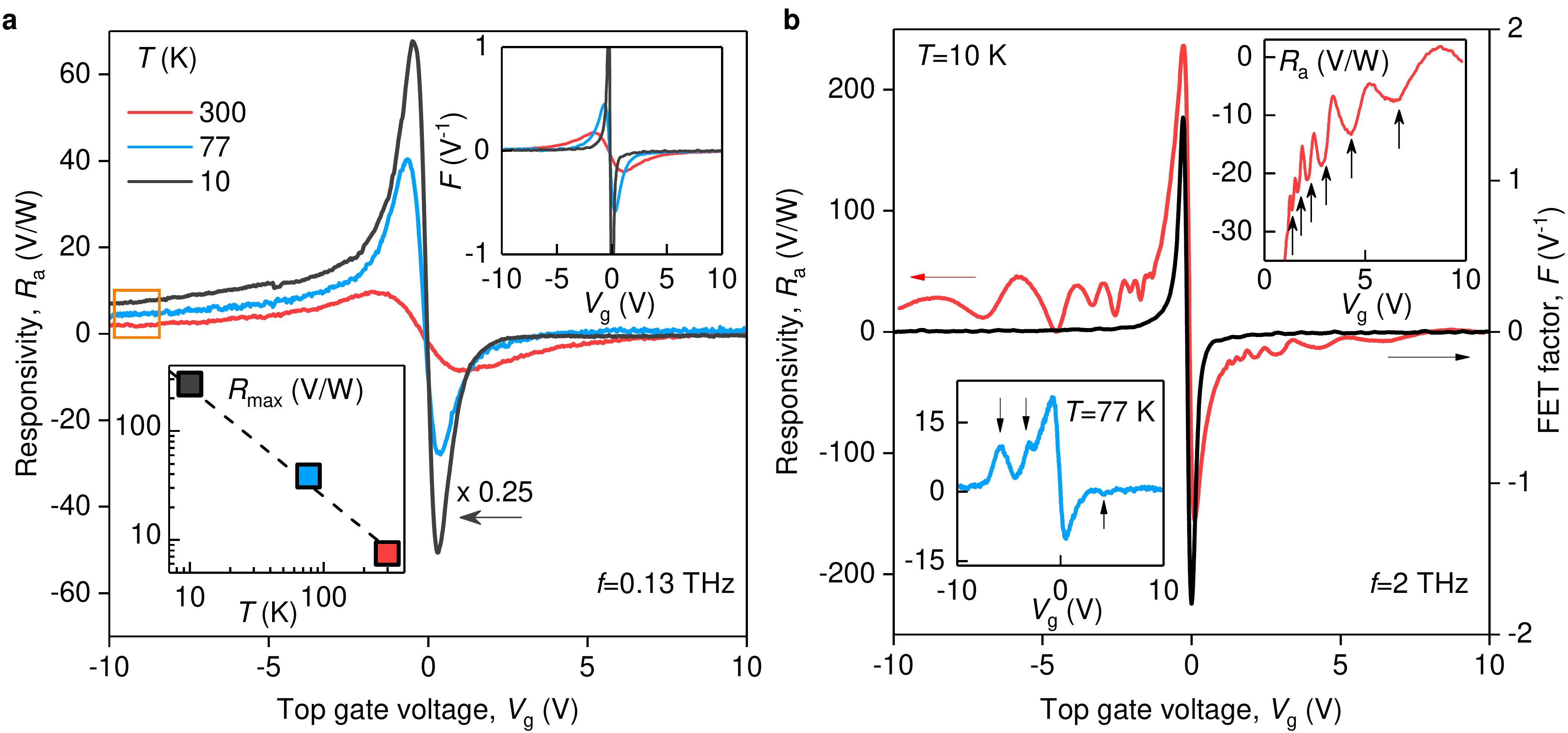}
	\caption{\textbf{Plasmon-assisted resonant THz photodetection.} \textbf{a,} Responsivity measured at $f=130$ GHz and three representative temperatures. Orange rectangle highlights an offset stemming from the rectification of incident radiation at the p-n junction between the p-doped graphene channel and the n-doped area near the contact. Upper inset: FET-factor $F$ as a function of $V\sous{g}$ at the same $T$. Lower inset: Maximum $R\sous{a}$ as a function of $T$. \textbf{b,} Gate dependence of responsivity recorded under $2~{\rm THz}$ radiation. The upper inset shows a zoomed-in region of the photovoltage for electron doping. Resonances are indicated by black arrows. Lower inset: resonant responsivity at liquid-nitrogen temperature.}
	\label{Fig_2}
\end{figure*}

\section*{RESULTS}
\subsection*{Graphene-based THz detectors}
There are three crucial steps to consider in the design of resonant photodetectors. First, the incoming radiation needs to be efficiently compressed into plasmons propagating in the FET channel. Second, the channel should act as a high-quality plasmonic cavity, where constructive interference of propagating plasma waves leads to the enhancement of the field strength. Third, the high-frequency plasmon field needs to be rectified into a dc photovoltage. To meet these hard-to-satisfy~\cite{Vicarelli2012, Spirito2014,Tong2015, Qin2017} requirements, we fabricated  proof-of-concept detectors using high-mobility bilayer graphene (BLG) FETs. To this end, we first applied a standard dry transfer technique to encapsulate BLG between two relatively thin ($d\approx80$ nm) slabs of hBN~\cite{Kretinin2014}. The heterostructure had side contacts (Fig.~\ref{Fig_1}a) which were extended to the millimeter scale and one of them served as a sleeve of the broadband antenna, Fig.~\ref{Fig_1}c and Supplementary Fig.~3a-b (see Methods). Another antenna sleeve was connected to the top gate covering the FET channel (inset of Fig. \ref{Fig_1}d). In this coupling geometry, the incident radiation induces high-frequency modulation of the gate-to-channel voltage thereby launching plasma waves from the source terminal~\cite{Dyakonov1996a}. The detector was assembled on a THz--transparent Si wafer attached to a Si lens focusing the incident radiation onto the antenna (Fig. \ref{Fig_1}b). 

We studied four BLG FETs, from $3$ to $6$ $\mu$m in length $L$ and from $6$ to $10$ $\mu$m in width $W$, all exhibiting typical field-effect behaviour as seen from measurements of the conductance $G$  (Fig.~\ref{Fig_1}d and Supplementary Fig.~3e). In particular, $G$ is minimal at the charge neutrality point and rises with increasing $V\sous{g}$. The mobility of our devices at the characteristic carrier density $n=10^{12}$ cm$^{-2}$ exceeded $10~{\rm m^2}/{\rm Vs}$ and remained above $2~{\rm m^2}/{\rm Vs}$ at temperatures $T=10$ K and $300$ K, respectively,  as determined from the characterization of a multiterminal Hall bar produced under identical protocol reported in Methods (Supplementary Note 1 and Supplementary Fig. 1).

\begin{figure*}[ht!]
	\centering\includegraphics[width=0.9\linewidth]{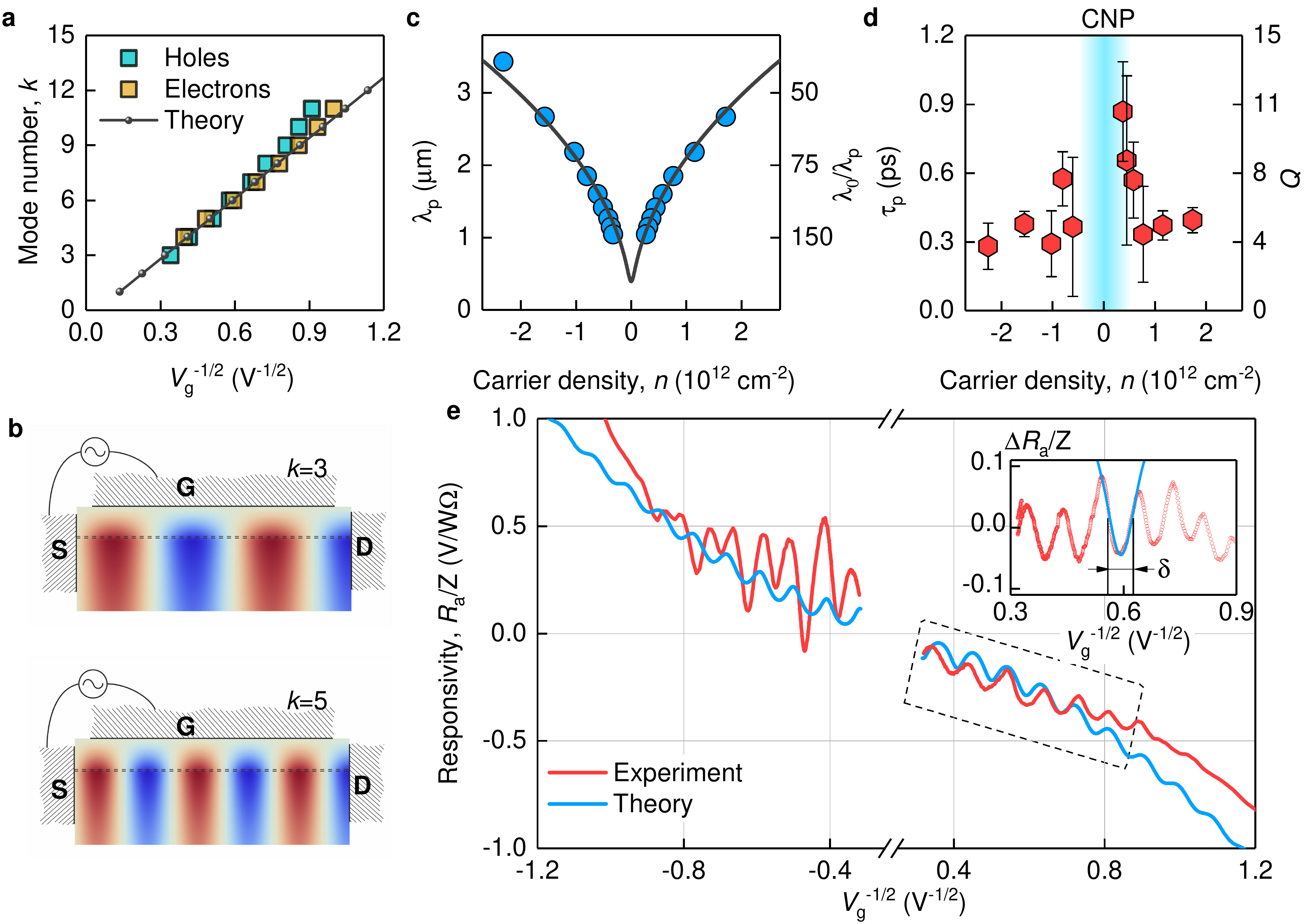}
	\caption{\textbf{Plasmon resonances in  encapsulated-graphene FET. }  \textbf{a,} Mode number $k$ as a function of $V\sous{g}^{-1/2}$ (symbols). Solid line: Theoretical dependence for $L=6~{\rm \mu m}$, $m=0.036m_e$  and $f=2~{\rm THz}$. The first mode supported by our Fabry-P\'{e}rot plasmonic cavity corresponds to $k_{\min} = 3$; the fundamental mode with $k=0$ is beyond the accessible gate voltages. \textbf{b,} Examples of high-frequency potential distribution in the plasmon mode (real part) under resonant conditions for given $k$. Brown and blue colours represent positive and negative values of electrical potential, respectively. S,G and D stand for source, gate, and drain terminals, respectively.  \textbf{c,} Experimental (symbols) and calculated (solid line) plasmon wavelengths $\lambda_{\rm p}$ as functions of carrier density, as obtained from (a). The corresponding value of the inverse compression ratio, $\lambda_{0}/\lambda_{\rm p}$, for $f=2~{\rm THz}$ is given on the right axis. \textbf{d,} Plasmon lifetime $\tau_{\rm p}$ and quality factor $Q$ as obtained from the width of the resonances shown in (e). Error bars stem from the fitting procedure. \textbf{e,} Experimental and calculated responsivities as functions of $V\sous{g}^{-1/2}$, normalized to the effective antenna impedance $Z=V\sous{a}^2/P$ relating the incident power to the resulting gate-to-channel voltage $V\sous{a}$. The theoretical Dyakonov-Shur dependence was obtained by using characteristic $\tau_{\rm p}=0.6~{\rm ps}$ from (d). Inset: normalized responsivity $R\sous{a}/Z$ after the subtraction of a smooth non-oscillating background. The solid blue line is the best Lorentzian fit to the data, with $\delta=0.1$ $V^{-1/2}$, which translates to $\tau_{\rm p}=0.5~{\rm ps}$.
	\label{Fig_3}}
\end{figure*}

\subsection*{Broadband operation }

We intentionally start the photoresponse measurements at the low end of the sub-THz domain, where the plasma oscillations are overdamped (see below). This allows us to compare the performance of our detectors with previous reports~\cite{Vicarelli2012, Spirito2014, LKoppens2014, Tong2015, Qin2017}. Figure~\ref{Fig_2}a shows an example of the responsivity $R\sous{a}=\Delta U/P$, where $\Delta U$ is the emerging source-to-drain photovoltage and $P$ is the incident radiation power, as a function of the top gate voltage $V\sous{g}$ under irradiation with frequency $f=0.13$ THz in one of our BLG detectors (see Methods). In good agreement with the previous studies, the $R\sous{a}(V\sous{g})$ dependence follows the evolution of the FET-factor $F=-\frac{1}{\sigma}\frac{d\sigma}{d V\sous{g}}$, shown in the inset of Fig.~\ref{Fig_2}a. In particular, $R\sous{a}$ increases in magnitude upon approaching the charge neutrality point (NP) where it flips sign because of the change in charge carrier type. 

We have studied the operation of our detectors at different temperatures and found that $R\sous{a}$ grows with decreasing $T$ (bottom inset of Fig. \ref{Fig_2}a) and reaches its maximum $R\sous{a}\approx240$~V/W at $T=10$~K due to a steeper $F(V\sous{g})$ at this $T$ (top inset of Fig. \ref{Fig_2}a). At large positive $V\sous{g}$, $R\sous{a}$ approaches zero at all $T$, whereas at negative $V\sous{g}$, a positive offset is observed (orange rectangle in Fig. \ref{Fig_2}a). This behaviour is common for this type of devices and is related to additional rectification by p-n junctions at the boundaries between the p-doped graphene channel and the n-doped contact regions~\cite{Cai2014, Ryzhii2006, Bandurin2017}.

The overall broadband responsivity of our BLG detectors is further improved in transistors with stronger nonlinearity, which can be conveniently parametrized by the FET-factor introduced above. To this end, we took advantage of the gate-tunable band structure of BLG and fabricated a dual-gated photodetector. Simultaneous action of the two gates results in a band gap opening and a steep $F(V\sous{g})$ dependence that, in turn, causes a drastic enhancement of $R\sous{a}$ (Supplementary Note 2). The latter exceeded 3 kV/W for a weak displacement field $D$ of 0.1 V/nm (Supplementary Fig. 2b). This translates to the  noise equivalent power (NEP) of about $0.2$~pW/Hz$^{1/2}$, estimated using the Johnson-Nyquist noise spectral density obtained for the same $D$. The observed performance of our detectors makes them competitive not only with other graphene-based THz detectors operating in the broadband regime~\cite{LKoppens2014}, but also with some commercial superconducting and semiconductor bolometers operating at the same $f$ and $T$ (Supplementary Table 1).  

\subsection*{Resonant photoresponse}

The response of our photodetectors changes drastically as the frequency of incident radiation is increased. Fig.~\ref{Fig_2}b shows the gate voltage dependence of $R\sous{a}$ recorded in response to $2$~THz radiation. In stark contrast to Fig.~\ref{Fig_2}a, $R\sous{a}$ exhibits prominent oscillations, despite the fact that $F$ as a function of $V\sous{g}$ is featureless (black curve in Fig. \ref{Fig_2}b). The oscillations are clearly visible for both electron and hole doping and display better contrast on the hole side, likely because of the aforementioned p-n junction rectification. Resonances are well discerned at $10~{\rm K}$, although they persist up to liquid-nitrogen $T$, especially for $V\sous{g}<0$. A further example of resonant operation of another BLG device is shown in Supplementary Note~3.

We have also studied the performance of our detectors at intermediate frequencies and found that the resonant operation of our devices onsets in the middle of the sub-THz domain  (Supplementary Note~4). In particular, we have found that at $f=460$ GHz, the resonances are already well-developed (Supplementary Fig.~4). At such low $f$, only two peaks in the photoresponse (one for electrons and one for holes) are observed for the same gate voltage span as in Fig. 2b along with an apparent increase of their full width at half-height. These observations are in full agreement with the plasmon-assisted photodetection model discussed below.

\subsection*{Plasmon resonances in graphene FETs}
We argue that the observed peaks in the photoresponse emerge as a result of plasmon resonance in the FET channel. To this end, we model our FET as a plasmonic Fabry-Perot cavity endowed with a rectifying element. This results in responsivity given by
\begin{equation}
\label{Resonant-responsivity}
R\sous{a} = \frac{R_0}{|1 - r\sous{s} r\sous{d} e^{2i q L}|^2},
\end{equation}
where $R_0$ is a smooth function of carrier density $n$ and frequency $f$ that depends on the microscopic rectification mechanism, $r\sous{s}$ and $r\sous{d}$ are the wave reflection coefficients from the source and drain terminals, respectively, and $q$ is the complex wave vector governing the wave propagation in the channel (Supplementary Note~5). In gated 2D electron systems, the relation between the frequency $\omega$ and the real part of the wave vector $q'$ is linear, $\omega=sq'$, where the plasmon phase velocity is
\begin{equation}
\label{Wave-velocity}
s =v\sous{F} \sqrt{4 \alpha_c k\sous{F} d} = \sqrt{ \frac{e}{m}|V\sous{g}|}.
\end{equation}
Here $m$ and $e$ are the effective mass of carriers and the elementary charge respectively, $v\sous{F}$ and $k\sous{F}=\sqrt{\pi n}$ are the Fermi velocity and the Fermi wave vector, $d$ is the distance to the gate, $\alpha_c = e^2/(4\pi\varepsilon\sous{z} \varepsilon_0 \hbar v_F)$ is the dimensionless coupling constant and $\varepsilon\sous{z}$ is the out-of-plane dielectric permittivity\cite{Chaplik1972,Tomadin2013}. We further note that eq.~(\ref{Wave-velocity}) is valid for monolayer graphene upon replacement of the effective mass $m$ with the cyclotron mass, $m\rightarrow \hbar k\sous{F}/v\sous{F}$ (see Supporting Note~6). The latter increases with gate-induced carrier density $n$, thereby limiting the tuning range of $s$ for a given voltage span. In contrast, in the case of BLG, $m$ is nearly constant ($\approx0.036m_e$) for experimentally accessible values of $V\sous{g}$, a feature that allows us to vary $s$ over a wider range and thus switch the detector between multiple modes, as we now proceed to show.

It follows from eq.~(\ref{Resonant-responsivity}), that the responsivity of our Fabry-Perot rectifier is expected to peak whenever the denominator in eq.~(\ref{Resonant-responsivity}) approaches zero. In our devices, the source potential is clamped to antenna voltage, and no ac current flows into the drain, therefore $r\sous{s} r\sous{d} \approx - 1$ (Ref.~\citen{Dyakonov1996a,Bandurin2017}). The resonances should therefore occur whenever the real part of the wave number is quantized according to
\begin{equation}
\label{q-quantization}
q' = \frac{\pi}{2L} (2k+1), \qquad k=0,\,1,\,2...
\end{equation}
The quantization rule (\ref{q-quantization}) combined with eq. (\ref{Wave-velocity}) predicts a linear dependence of the mode number $k$ on $|V\sous{g}|^{-1/2}$ which may serve as a benchmark for plasmon resonances in the FET channel. This is indeed the case of our photodetector, as shown in Figs. \ref{Fig_3}a,e and Supplementary Fig.~3c. The slope of the experimental $k(|V\sous{g}|^{-1/2})$ dependence in Fig.~\ref{Fig_3}a matches well the theoretical expectation for a BLG Fabry-Perot cavity of length $L=6$~$\mu$m. At large $|V\sous{g}|^{-1/2}$, we find a slight upward trend in the experimental data with respect to the linear dependence. We attribute this trend to deviations of the plasmon dispersion from the linear law at short wavelengths which stem from the non-local relation between electric potential and carrier density~\cite{Chaplik1972}. Note that the known non-parabolicity of the BLG spectrum~\cite{Zou2011} resulting in an increase of $m$ at large density $n$ would bend the dependence in Fig.~\ref{Fig_3}a in the opposite direction.

\subsection*{Photovoltage-based spectroscopy of 2D plasmons}
The resonant gate-tunable response of our detectors offers a convenient tool to characterize plasmon modes in graphene channels. From eq.~(\ref{q-quantization}) it follows that resonances occur if $L = (2k+1) \lambda_{\rm p}/4$, where $\lambda_{\rm p}=2\pi/q'$ is the plasmon wavelength (Fig.~\ref{Fig_3}b). Using the experimentally observed peak positions, we have determined the density dependence of $\lambda_{\rm p}$, shown in Fig.~ \ref{Fig_3}c, which flaunts excellent agreement with theory. The compression ratio $\lambda_{\rm p}/\lambda_{0}$ between the plasmon and free-space wavelength ($\lambda_{0} = c/f$ and $c$ the speed of light in vacuum) ranges between $1/50$ and $1/150$, highlighting the ultra-strong confinement of THz fields enabled by graphene plasmons, matching the record value known in the literature~\cite{Woessner2015}. 

Apart from ${\lambda\sous{p}}$, the resonant responsivity carries information about another valuable characteristic of plasmons, namely, their lifetime, $\tau_{\rm p}$. The latter is related to the peak width at half-height $\delta$ via (Supplementary Note~7)
\begin{equation}
V\sous{g}^{-1/2} / \delta = \omega \tau_{\rm p}~.
\end{equation}
Using Lorentzian fits to the photoresponse curves (inset of Fig.~\ref{Fig_3}e), we have extracted $\tau_{\rm p}$ as a function of $n$, shown in Fig.~\ref{Fig_3}d. The lifetime was found to range between $\approx 0.3$ and $\approx 0.9~{\rm ps}$, which is slightly shorter than the transport time $\tau_{\rm tr}\approx 2~{\rm ps}$ as extracted from the mobility, $\tau_{\rm tr}=m\mu/e$ (Supplementary Note~1). The corresponding quality factor, $Q=2\pi f \tau_{\rm p}$, was found to vary between $4$ and $11$ for $f=2~{\rm THz}$, and between $0.2$ and $0.7$ for $f=0.13~{\rm THz}$, see Fig.~\ref{Fig_3}d. The latter implies that it is unreasonable to expect resonant photoresponse of such detectors in the GHz range, and they can only operate in the broadband (non-resonant) regime, in accordance with the data in Fig. \ref{Fig_2}a. On the contrary, the resonant responsivity should become more profound at higher frequencies of the THz window and can be further enhanced in graphene FETs of higher quality, such as those using graphite gates to screen remote charge impurities~\cite{AFY2018}. 

\begin{figure*}[ht!]
	\centering\includegraphics[width=0.5\linewidth]{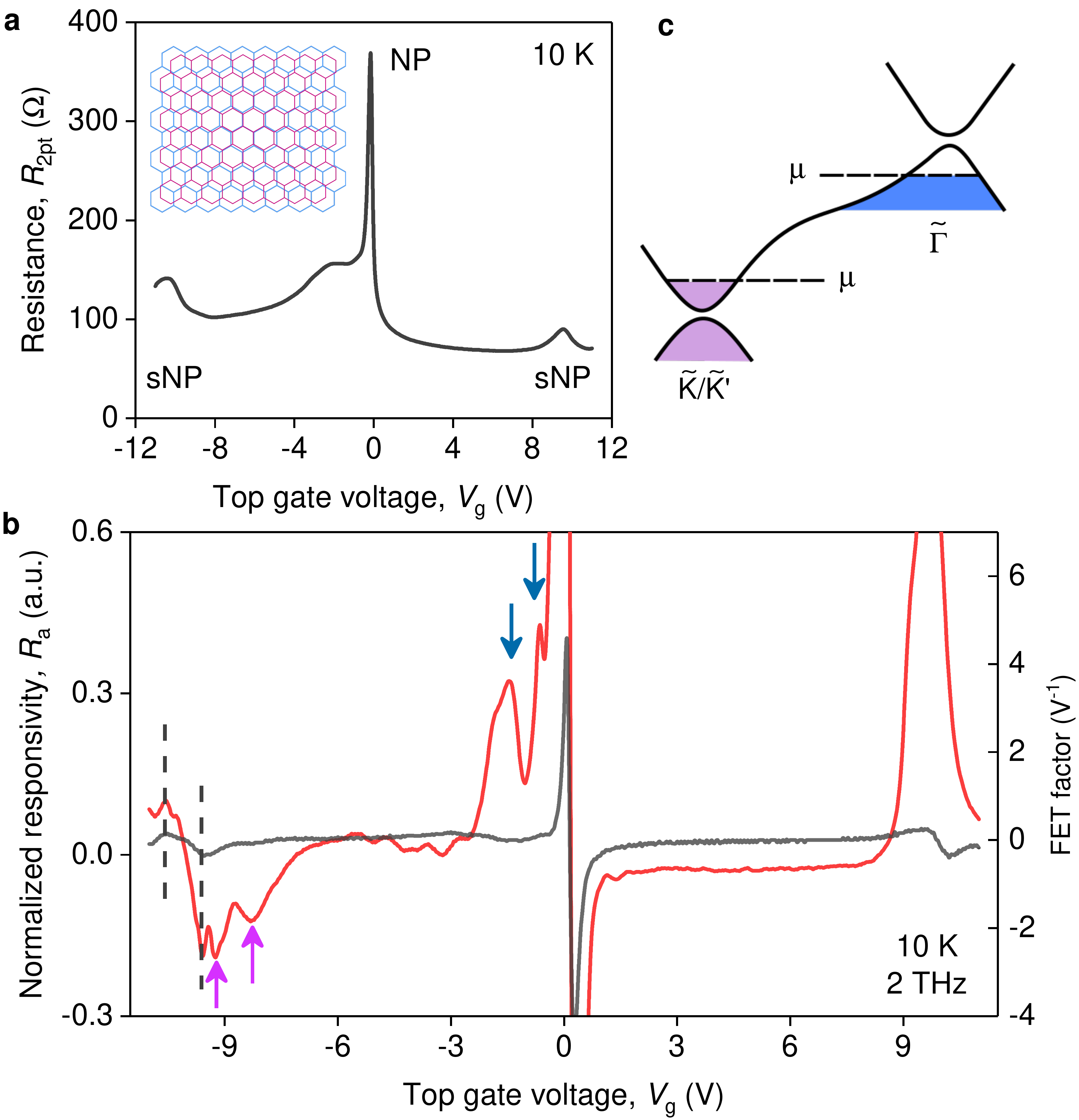}
	\caption{\textbf{Miniband plasmons in BLG/hBN moir\'e superlattices}. \textbf{a,} Two-terminal resistance of one of our BLG/hBN superlattice devices as a function of $V\sous{g}$  measured at given $T$. Inset: Illustration of the BLG/hBN superlattice demonstrating a mismatch between graphene and hBN lattice constants. For simplicity, only one graphene layer is shown. \textbf{b,} Normalized responsivity (red) and the FET-factor (black) as a function of $V\sous{g}$ measured in the same device as in (a). Dashed line`s trace $V\sous{g}$ where the FET-factor reaches extreme values in the vicinity of the sNP. Pink (blue) arrows point to the resonant peaks near the secondary (main) NP. $L=3$ $\mu$m. \textbf{c,} Schematic representation of the BLG/hBN superlattice band structure. In the vicinity of the $\tilde\Gamma$-point (blue), BLG supports propagation of the ordinary plasma waves. Miniband THz plasmons emerge when the chemical potential approaches the sNP (pink).}
	\label{Fig:superlattice}
\end{figure*}

\subsection*{Miniband plasmons in graphene/hBN superlattices}

The  approach demonstrated above is universal and can be applied to studies of plasmons in arbitrary high-mobility 2D systems embedded in FET channels, as we now proceed to show for the case of devices made of BLG/hBN moir\'e superlattices~\cite{Mucha-Kruczyński2013}.

Figures \ref{Fig:superlattice}b and Supplementary Fig. 3c show examples of $R\sous{a}$ as a function of $V\sous{g}$ recorded in our superlattice devices in response to 2 THz radiation. As in the case of plain BLG, the overall evolution of the superlattice responsivity $R\sous{a}(V\sous{g})$ follows that of the FET-factor (black curve) modulated by the plasmon resonances. Note the total number of resonances is smaller due to the shorter FET channel (cf. Supplementary Fig. 3c) and they are visible only for $V\sous{g}<0$, presumably due to a stronger nonlinearity in this detector for negative doping (in another superlattice FET, the resonances were well-observed for both $V\sous{g}$ polarities as shown in Supplementary Fig. 3c). Importantly, the FET-factor in these devices is, in turn, a more complex function of $V\sous{g}$ (cf. inset of Fig. 2a) due the presence of secondary neutrality points (sNP) stemming from a peculiar band structure of the BLG/hBN superlattice. The latter is characterized by narrow minibands emerging in the vicinity of the $\tilde K/ \tilde K'$-points of the superlattice Brillouin zone~\cite{Mucha-Kruczyński2013} (Fig. \ref{Fig:superlattice}c). The sNPs are clearly visible as peaks in the FET resistance which appear around $V\sous{g}=\pm10$ V (Fig. \ref{Fig:superlattice}a).

A striking feature of the superlattice photoresponse is the resonances appearing when the Fermi level is brought close to the sNP (pink arrows in Fig. \ref{Fig:superlattice}b). The resonances are of opposite sign with respect to those observed near the main NP (blue arrows), which indicates that they originate from the plasmons supported by the charge carriers of the opposite type (cf. Fig. 2b). Since the latter are hosted by the minibands near the $\tilde K/ \tilde K'$-points of the superlattice Brillouin zone (Fig. \ref{Fig:superlattice}c), our measurements provide evidence for miniband plasmons that were long identified theoretically~\cite{Tomadin2014} but remained elusive in experiment. To date, the experimental studies of superlattice plasmons have been only performed at room temperature using scattering-type scanning near field microscopy operating in the mid-IR domain~\cite{Ni2015}. The mid-IR excitation energy (10 $\mu$m $\approx120$ meV) is high enough to induce interband absorption close to the sNP, which hampers the observation of plasmons in superlattice minibands~\citen{Ni2015}. In contrast, our approach relies on the low-energy excitations (2 THz $\approx8$ meV), is applicable at cryogenic temperatures, and, therefore, paves a convenient way for further studies of miniband plasmonics.

\section*{DISCUSSION}

Resonant responsivity is a universal phenomenon in ultra-clean graphene devices and is expected to be independent of the physical mechanisms behind the rectification of the ac field into a dc photovoltage. Nevertheless, it is important to establish possible nonlinearities responsible for the rectification, for example, in order to be able to increase the magnitude of responsivity. 

We first note that the aforementioned asymmetry in $R\sous{a}(V\sous{g})$ between electron and hole doping indicates rectification at the $p$-$n$ junction formed in vicinity of the contacts. This rectification usually appears due to the  thermoelectric effect arising as a result of non-uniform sample heating and the difference between the Seebeck coefficients in the graphene channel and contact regions\cite{Cai2014,Alonso-Gonzalez2016,Jung2016,Bandurin2017} (Supplementary Note~8). However, $R\sous{a}$ remains finite even for $V\sous{g}>0$, where both channel and contact areas are $n$-doped. This indicates that alternative rectification mechanisms are also involved.

Another commonly accepted mechanism is the rectification arising as a result of the simultaneous action of longitudinal high-frequency field and modulation of channel conductivity, also known as resistive self-mixing~\cite{Daryoosh2013}. The latter can be enhanced by the dc photovoltage that balances the difference between electron kinetic energies at the source and drain terminals~\cite{Dyakonov1996a,Daryoosh2013}, similar to  Bernoulli's law for classical fluids. Both mechanisms are combined into so-called Dyakonov- Shur (DS) rectification~\cite{Dyakonov1996a} and result in $R\sous{0}$ proportional to the sensitivity of the conductivity to the gate voltage variation~\cite{Knap2009}, given by the $F$-factor introduced above (Supplementary Note~9). In Fig.~\ref{Fig_3}e we compare the resonant photoresponse of our BLG photodetector with the responsivity expected from the DS model~\cite{Dyakonov1996a} assuming an average $\tau \sous{p}\sim0.6$ ps, as found from Fig. \ref{Fig_3}d, and using the effective antenna impedance $Z$ as the only fitting parameter. The two curves show the same functional behaviour and match quantitatively for the n-doped case (where the p-n junction is absent) and $Z\approx 74$~$\Omega$, a value close to that expected from the equivalent circuit design~\cite{Bandurin2017}. We further note, that although the original DS proposal was based on the hydrodynamic electron transport~\cite{BandurinFO201,Berdyugin2018}, an identical photoresponse is expected outside the hydrodynamic window as it follows from the analysis of graphene's nonlinear conductivity~\cite{Alepr2018}.  
 
Last but not least, we note that while the overall trend of the responsivity is well-described by the model introduced above, the values of $\tau\sous{p}$ extracted from the peak width at half-height are found to be below the momentum relaxation time. This suggests that other mechanisms of resonance broadening are also involved. In particular, leakage of plasma waves into metal contacts~\cite{Satou2009} and electromagnetic dissipation in antenna may also contribute to the apparent resonance width. We have found that respective contributions to $\tau\sous{p}^{-1}$ are most pronounced at large carrier densities and small harmonic numbers (Supplementary Figs. 6 and 7), in agreement with experimental data in Fig. 3d. Elimination of these damping channels, e.g. with Schottky/tunnel contacts and low-impedance antennas, may extend the resonant detection down to tens of gigahertz~\cite{Graef2018}. Other dissipation channels such as electron viscosity~\cite{Kumar2017,Svintsov2017} and interband absorption~\cite{Chen2012} should be most pronounced at higher-order harmonics and in the vicinity of the NP, as opposed to the data in Fig. 3d, and are unlikely relevant to the present study. 

In conclusion, we have shown that high-mobility graphene FETs exploiting far-field coupling to incoming radiation can operate as resonant THz photodetectors. In addition to their potential applications in high-responsivity detection and on-chip spectroscopy of the THz radiation, our devices also represent a convenient tool to study plasmons under conditions where other approaches may be technically challenging. Due to their compact size and far-field coupling, our photodetectors can easily be employed to carry out plasmonic experiments in extreme cryogenic environments and in strong magnetic fields, as well in studies of more complex van der Waals heterostructures. As an example, we have demonstrated the use of our approach to reveal low-energy plasmons hosted by moir\'e minibands in BLG/hBN superlattices. The method has a strong potential for studies of collective modes in magnetic minibands which have recently gained a great level of attention~\cite{KrishnaKumar2017}.
\newline

\section*{METHODS}

\subsection*{Device fabrication}
All our devices were made of BLG. BLG was first encapsulated between relatively thick hBN crystals using the standard dry-peel technique \cite{Kretinin2014}. The thickness of the top hBN was measured by atomic force microscopy. The stack was then deposited either directly on top of a low-conductivity boron-doped silicon wafer capped with a thin oxide layer ($500$ nm) or on a predefined back gate electrode. The resulting van der Waals heterostructure was patterned using electron beam lithography to define contact regions. Reactive ion etching was then used to selectively remove the areas unprotected by a lithographic mask, resulting in trenches for depositing electrical leads. Metal contacts to graphene were made by evaporating 3 nm of Cr and 60 nm of Au. Afterwards, a second e-beam lithography was used to design the top gate. The graphene channel was finally defined by a third round of e-beam lithography, followed by reactive ion etching etching using Poly(methyl methacrylate) and gold top gate as the etching mask. Finally, we used optical photolithography to pattern large antenna (spiral or bow-tie) sleeves connected to the source and the top-gate terminals, followed by evaporation of 3 nm of Cr and 400 nm of Au. Antennas were designed to operate at an experimentally accessible frequency range.

\subsection*{Photoresponse measurements}
Photoresponse measurements were performed in a variable temperature optical cryostat equipped with a polyethylene window that allowed us to couple the photodetector to incident THz radiation. The latter was focused to the device antenna by a silicon hemispherical lens attached the silicon side of the chip (Fig. \ref{Fig_1}b). The transparency of the chips to THz radiation over the entire temperature and frequency range was verified in transmission experiments using a home-made optical cryostat coupled to the THz spectrometer. Photovoltage measurements were performed using either a standard lockin amplifier synchronized with a chopper rotating at 1 kHz frequency, positioned between the radiation source and the cryostat window, or by a home-made measurement board. 

In order to study the photoresponse of our detectors at different frequencies, we used three radiation sources. Sub-THz radiation was provided by two backward wave oscillators (BWO) generating $f=0.13$ THz and $f=0.46$ THz. For higher frequencies, a quantum cascade continues wave laser based on a GaAs/Al$_{0.1}$Ga$_{0.9}$As heterostructure emitting $f=2.026$ THz radiation was used. 

The responsivity of our devices was calculated assuming that the full power delivered to the device antenna funnelled into the FET channel. The as-determined value provides the lower bound for our detectors' responsivity and is usually referred to as extrinsic. The calculation procedure consisted of a few steps. First the source-to-drain voltage $U\sous{dark}$ was measured as a function of $V\sous{g}$ in the dark. Then, the dependence of the source-to-drain voltage $U\sous{SD}$ on $V\sous{g}$ was recorded under illumination with THz radiation. The difference $\Delta U= U\sous{SD}-U\sous{dark}$ formed the photovoltage. At the next stage, we measured the full power $P\sous{full}$ delivered to the cryostat window using Golay cell. The responsivity was then calculated as $R\sous{a}=\Delta U/ P$, where $P\approx P\sous{full}/3.5$ is the power delivered to the device antenna after accounting for losses in the silicon lens and the cryostat optical window ($\approx5.5$ dB). All the measurements reported above were performed in the linear-in-$P$ regime. The performance of our detectors outside the linear regime is discussed in Supplementary Note~10 and reported in Supplementary Fig.~8.

\section*{Data availability}
The data that support the findings of this study are available from the corresponding author upon reasonable request.

\section*{Acknowledgements}
Device fabrication and Manchester’s part of the work was supported by the European Research Council, the Graphene Flagship and Lloyd’s Register Foundation. The work at the MSPU (Photoresponse measurements) has been carried out with the support of the Russian Science Foundation (project No. 17-72-30036). D.A.B. acknowledges financial support from Leverhulme Trust. Experimental work of M.M. (transport measurements) was supported by Russian Science Foundation (Grant 18-72-00234). M.P. is supported by the European Union’s Horizon 2020 research and innovation programme under grant agreement No. 785219 - GrapheneCore2. Modelling of antenna electrodynamics was supported by RFBR (Project 18-29-20116). Theoretical work of D.S. was supported by the grant 16-19-10557 of the Russian Science Foundation. Photoresponse measurements have been performed using quantum cascade laser fabricated by A. Valavanis in the group of Prof. Dragan Indjin in the University of Leeds (UK). We thank A. Tomadin, R. Krishna Kumar, A. Berdyugin, L. Levitov and V. Fal'ko for fruitful discussions.

\section*{Author contributions}
D.A.B. and G.F. designed and supervised the project. S.G.X. and I.G. fabricated the devices. Photoresponse measurements were carried out by I.G., M.M and D.A.B. Data analysis was performed by D.A.B. and D.S. Theory analysis was done by D.S. The manuscript was written by D.A.B. and D.S. with input from I.V.G., M.P., A.P. and A.K.G. Experimental support was provided by I.T., D. Y., S.Z. and G.G. T.T. and K.W. grew the hBN crystals. All authors contributed to discussions.

\section*{Competing interests}
The authors declare no competing interests.

\newpage
\clearpage

\onecolumngrid
\setcounter{equation}{0}
\setcounter{figure}{0}
\renewcommand{\thesection}{}
\renewcommand{\thesubsection}{\arabic{subsection}}
\renewcommand{\theequation} {\arabic{equation}}
\renewcommand{\thefigure} {\arabic{figure}}
\renewcommand{\thetable} {\arabic{table}}

\section*{ }

\subsection*{Supplementary Note 1: Device characterization} 
\label{Section: characterization}
\hspace{1em}Our photodetectors represent two-terminal field-effect transistors (FET) and, therefore, the measured conductance (Fig. \ref{Fig_1}d of the main text), which contains non-zero contribution from the BLG-metal contact, does not provide the information on the quality of the FET channel. In order to estimate the mobility of charge carriers in the BLG channel, we fabricated a reference multi-terminal Hall bar using the same procedure as described in Methods. The Hall bar was characterized using the standard four-terminal geometry that involved the measurements of its sheet resistance $\rho$ as a function of carrier density $n$ and temperature $T$ (Supplementary Fig. 1). One can see a typical field-effect behavior for high-quality graphene that manifests itself in sharp peak in $\rho$ at the charge neutrality point which decays steeply with increasing $n$. The charge carrier mobility $\mu$ was calculated using the Drude formula, $\mu=\sigma/ne$, and for typical $n=10^{12}$ cm$^{-2}$ exceeded 10 m$^{2}$/Vs at liquid helium $T$ and remained around 2.5 m$^{2}$/Vs at room temperature. 

\begin{figure}[ht!]
	\centering\includegraphics[width=0.4\linewidth]{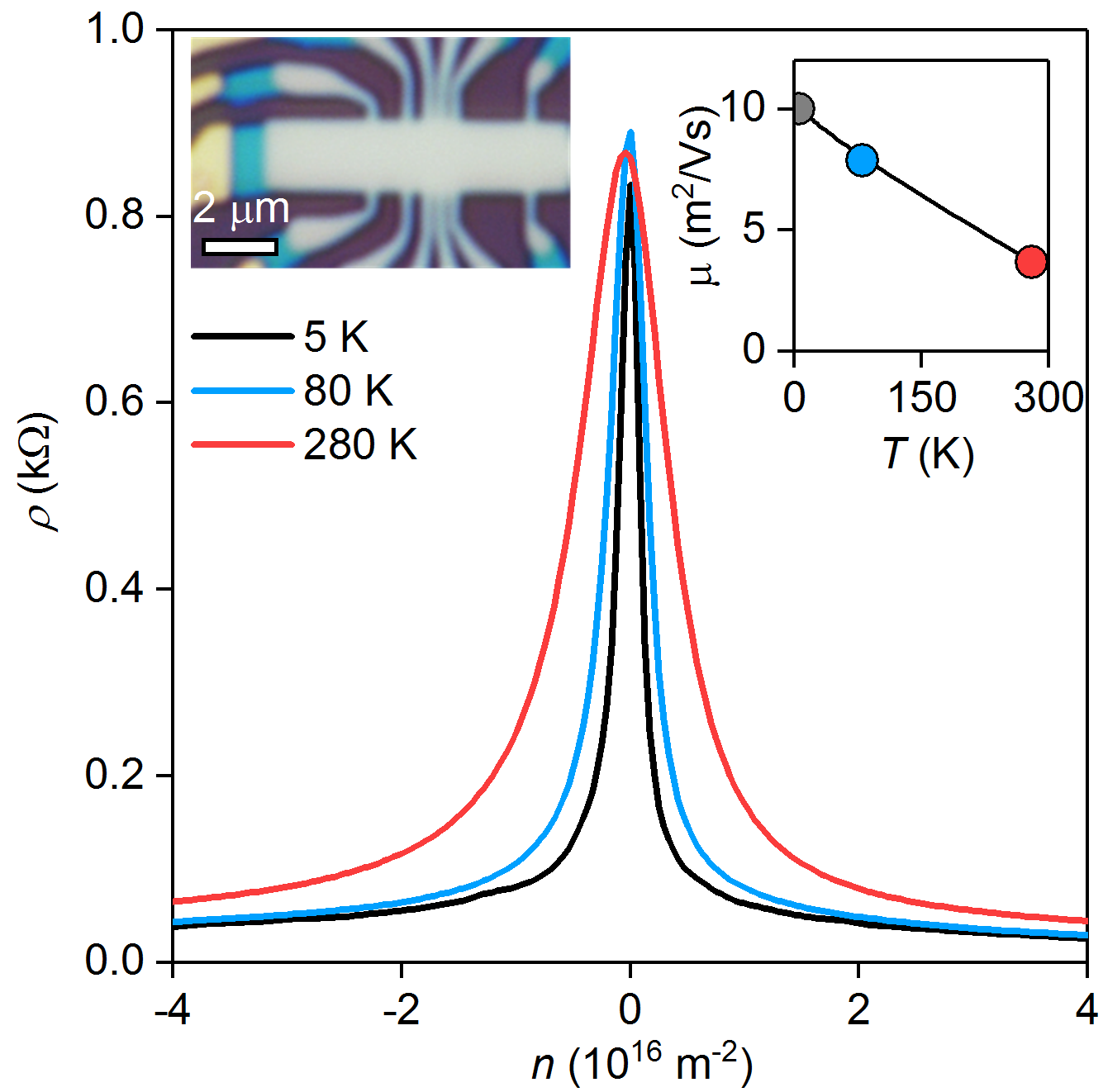}
	\caption{\textbf{Reference multiterminal BLG field effect transistor.} Sheet resistance as a function of $n$ for different $T$ measured in the standard four-terminal geometry. Left inset: Optical photographs of our reference Hall bar. Right inset: Mobility as a function of $T$ measured at $n=10^{12}$ cm$^{-2}$. }
	\label{Fig:Reference}
\end{figure}

\subsection*{Supplementary Note 2: High-responsivity THz detection in dual-gated BLG field effect transistors} 
\label{Section: R enhancement}
\hspace{1em}As discussed in the main text, the responsivity of the THz detectors made of the field effect transistors is proportional to the sensitivity of the FET conductivity to the gate voltage variation. To improve the performance of our detectors, we took advantage of BLG's gate-tunable band structure and fabricated a dual-gated device (top and bottom insets of Supplementary Fig. \ref{Fig:Gap_opening}a). The idea is that when an electric field is applied perpendicular to the channel it induces a band gap in BLG that leads to a steeper dependence of the FET resistance $R\sous{2pt}$ on the gate voltage. Supplementary Fig. \ref{Fig:Gap_opening}a shows examples of $R\sous{2pt}(V\sous{tg})$ dependences measured at few $V\sous{bg}$ demonstrating the expected increase of $R\sous{2pt}$ with increasing the average displacement field $D=\frac{\varepsilon}{2}(V\sous{tg}/d\sous{bg}-V\sous{bg}/d\sous{tg})$ applied to BLG, where $d\sous{bg}$  ($d\sous{bg}$) is the thickness of the bottom (top) hBN crystal and $\varepsilon$ is its dielectric constant. 

Supplementary Fig. \ref{Fig:Gap_opening}b shows the top gate voltage dependence of $R\sous{a}$ measured in response to 0.13 THz radiation in the dual-gated detector. In the case of zero back gate voltage (black curve), $R\sous{a}(V\sous{tg})$ repeats the behaviour of another detector reported in Fig. 2a of the main text. Namely, the responsivity reaches its maximum of about 200 V/W near the NP where it flips its sign because of the change in the charge carrier type. Note, the absolute value of the maximum responsivity is very close to that reported in the main text (Fig. 2b) highlighting the reproducibility of our detectors' performance. When the back gate voltage is applied, the responsivity increases drastically (red and blue curves in Supplementary Fig. \ref{Fig:Gap_opening}b). Already for a moderate $D\sim0.1$ V/nm the responsivity increases by more than an order of magnitude and exceeds 3 kV/W. The corresponding noise equivalent power, estimated using the Johnson-Nyquist noise spectral density for the same $D$, reaches 0.2 pW/Hz$^{1/2}$. This makes our detector competitive not only with other graphene-based THz detectors~\cite{LKoppens2014} but also with some commercial semiconductor and superconductor bolometers (Table~\ref{Table:THz_detectors}).

\begin{figure}[ht!]
	\centering\includegraphics[width=0.8\linewidth]{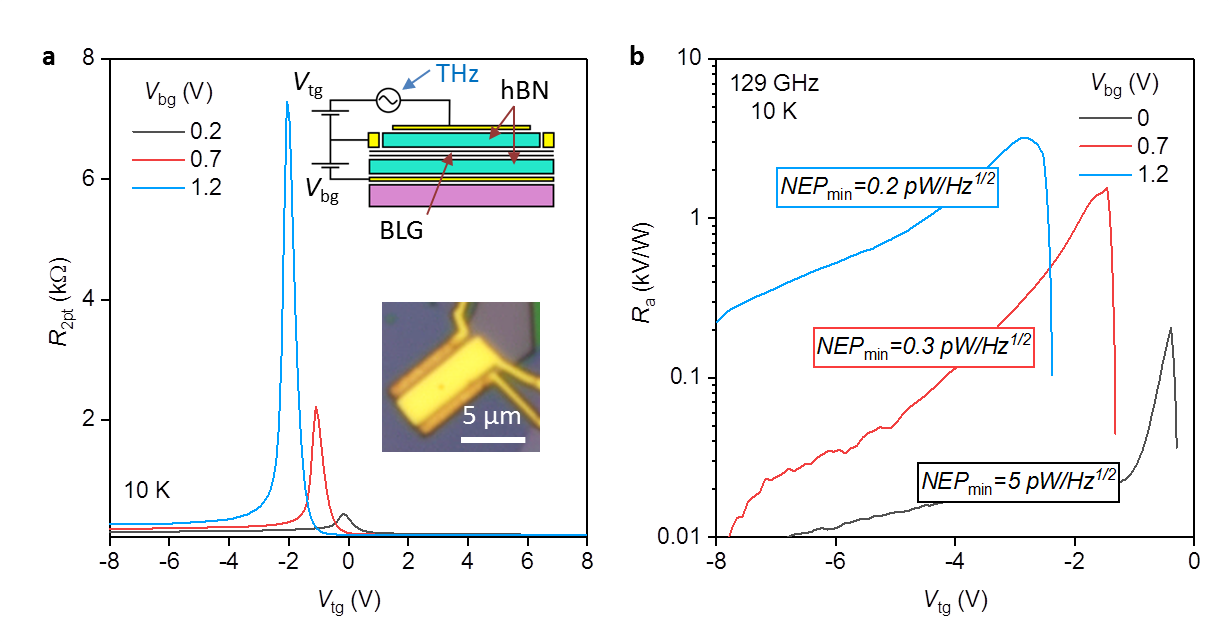}
	\caption{\textbf{Photoresponse of a dual-gated BLG detector}. \textbf{a,} Two-terminal resistance as a function of $V\sous{tg}$ measured in a dual-gated BLG FET for different $V\sous{bg}$. Top inset: Schematic of a dual-gated THz detector. Bottom inset: Optical photographs of the device. \textbf{b,} Responsivity as a function of $V\sous{tg}$ for different $V\sous{bg}$ measured at given $f$ and $T$. }
	\label{Fig:Gap_opening}
\end{figure}

\begin{table}[h!]
	\begin{tabular}{c|c|c|cl}
		\textbf{Detector}                                                                   & \textbf{NEP,  pW/Hz$^{0.5}$} & \textbf{Operation Temperature, K} & \textbf{Reference}                                                       &  \\ \cline{1-4}
		\begin{tabular}[c]{@{}c@{}}Superconducting \\ hot electron bolometer*\end{tabular}   & 0.1 - 1                 & 2.5 - 4.5                         & \begin{tabular}[c]{@{}c@{}}www.boselec.com\\ www.scontel.ru\end{tabular} &  \\ \cline{1-4}
		\begin{tabular}[c]{@{}c@{}}Semiconductor\\  hot electron bolometer (e.g. InSb)\end{tabular}   & 0.04 - 0.8              & 1.6 - 4.2                         & www.infraredlaboratories.com                                             &  \\ \cline{1-4}
		\begin{tabular}[c]{@{}c@{}}Dual-gated bilayer graphene \\ THz detector\end{tabular} & 0.2                     & 10                                & This work                                                                & 
	\end{tabular}
	\caption{\textbf{Comparison of cryogenic THz detectors}. *Reported values of NEP were taken for the same frequency (0.13 THz) used to probe our dual-gated devices.}
	\label{Table:THz_detectors}
\end{table}

\subsection*{Supplementary Note 3: Further examples of resonant photoresponse} \label{Section: Furhter examples}
\begin{figure}[ht!]
	\centering\includegraphics[width=0.75\linewidth]{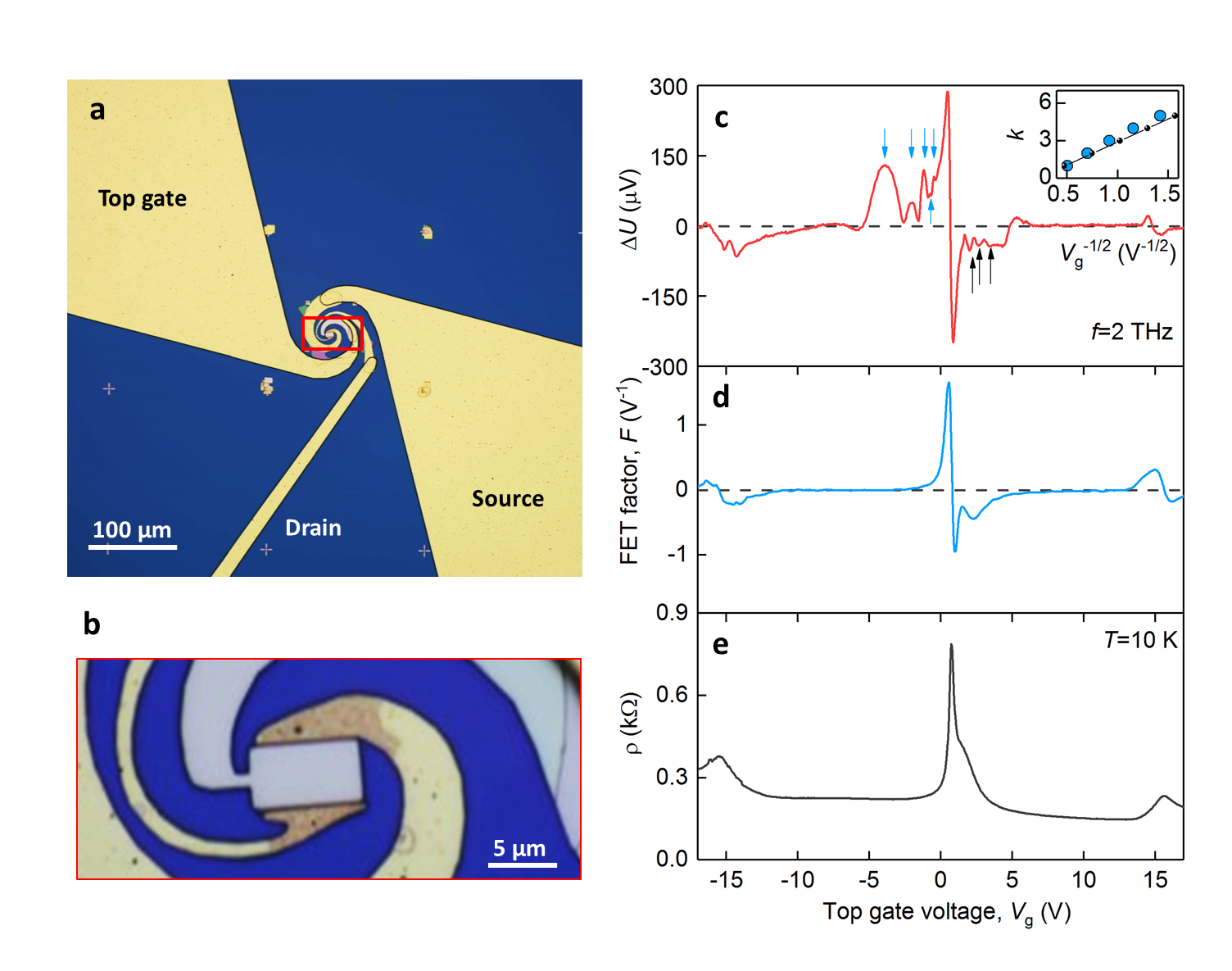}
	\caption{\textbf{Further examples of resonant photoresponse}. \textbf{a-b,} Optical photographs of another THz photodetector. FET channel is coupled to incoming radiation by a broadband logarithmic spiral antenna. The red rectangle in (a) indicates the region shown in (b). \textbf{c,} Photovoltage versus $V\sous{g}$ recorded as a response to $f=2$ THz radiation in one of our BLG/hBN superlattice devices. Arrows point to the resonant peaks. Inset: Mode number $k$ as a function of $V\sous{g}^{-1/2}$ taken from the peaks marked by the blue arrows. Black: theoretical dependence expected for $L=4$ $\mu$m, $m=0.036me$ and $r\sous{s}r\sous{d}=-1$.  \textbf{d,} FET-factor $F$ as a function of $V\sous{g}$ obtained from the data in (a). \textbf{e,} Resistivity as a function of $V\sous{g}$ for the device in (a-b) measured at $T=10$ K. Three peaks correspond to the secondary neutrality points of BLG/hBN superlattice. }
	\label{Fig:Aligned}
\end{figure}

\hspace{1em}To illustrate that the observed resonant photoresponse is reproducible for different electronic systems embedded in FETs of various lengths $L$ and coupled to different antennas, Supplementary Fig. \ref{Fig:Aligned}c shows another example of the photovoltage $\Delta U(V\sous{g})$ emerging when the incoming 2 THz radiation is coupled to the broadband logarithmic spiral antenna connected to another FET. The latter is made of BLG having its crystallographic axis aligned with those of hBN, that reveals itself in peculiar three-peaks $R(V\sous{g})$ structure, shown in Supplementary Fig. \ref{Fig:Aligned}e. The photoresponse curves are rather similar to those shown in Supplementary Fig. \ref{Fig_2}b of the main text, namely they follow the envelope trend set by the FET-factor $F=-\frac{1}{\sigma}\frac{d \sigma}{d V\sous{g}}$ (Supplementary Fig. \ref{Fig:Aligned}d) superimposed with the resonant peaks. The resonances are periodic in $V\sous{g}^{-1/2}$ (inset to Supplementary Fig. \ref{Fig:Aligned}c) and are clearly seen for both electron and hole  doping. Importantly, on the contrary to Fig.~\ref{Fig_2}b, the photoresponse now changes sign multiple times following non-trivial $F(V\sous{g})$ evolution.

%


\newpage
\afterpage{\clearpage}
\subsection*{Supplementary Note 4: Resonant detection of sub-THz radiation} \label{Section: Frequency dependence}
\begin{figure}[ht!]
	\centering\includegraphics[width=0.4\linewidth]{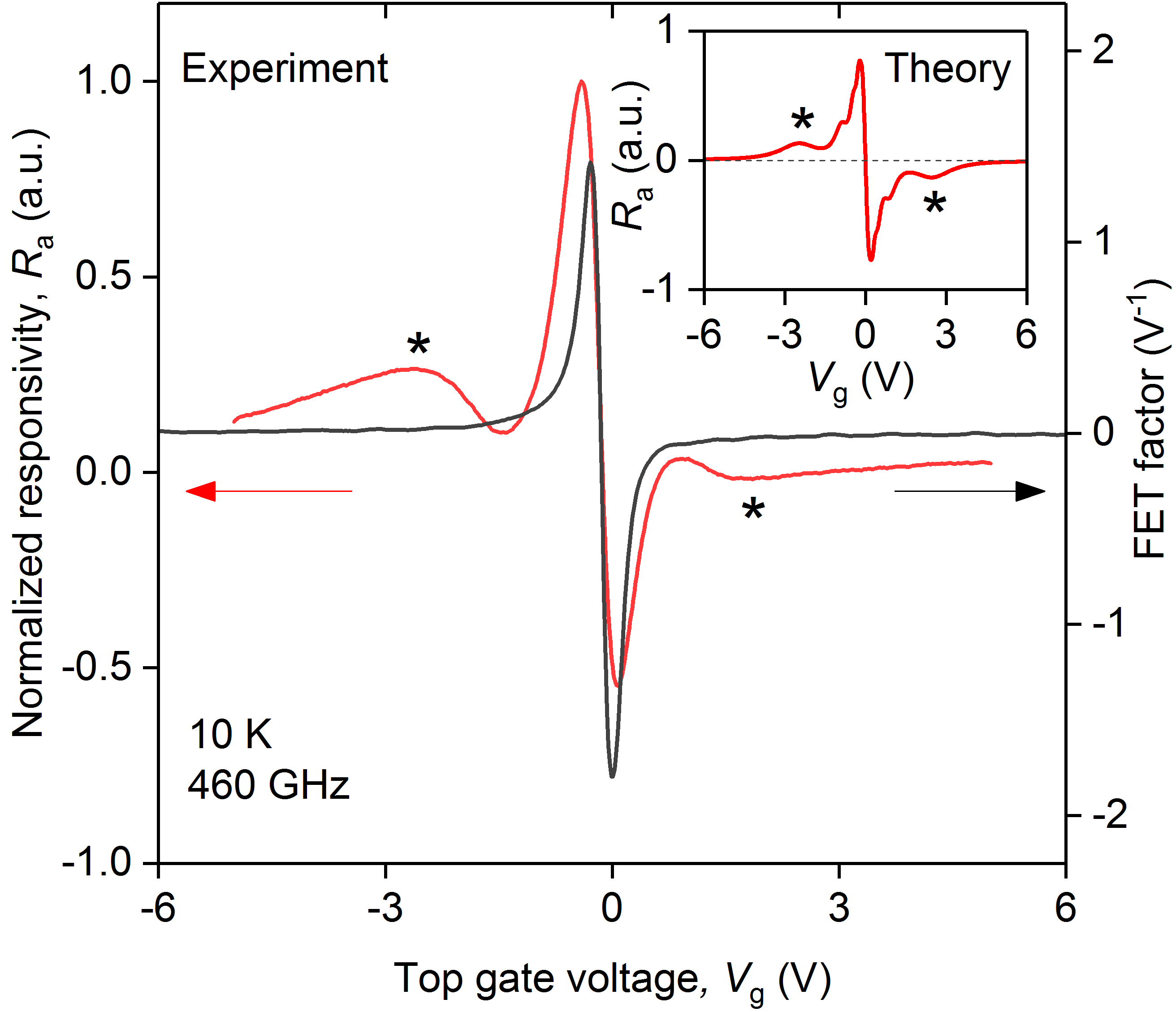}
	\caption{\textbf{Resonant detection in the sub-THz domain}. Normalized to unity $R\sous{a}$ as a function of $V\sous{g}$ measured in response to 460 GHz radiation. Data is acquired on the same device as in Fig. 2 of the main text. Two resonances are clearly visible for electron and hole doping and marked by the black stars. Black curve: FET-factor as a function of $V\sous{tg}$. Inset: Theory for $L=6$ $\mu$m, $f=460$ GHz, $m=0.036m\sous{e}$, $T=10$ K and $\tau=0.6$ ps (eq. \ref{DS_from_the_paper}). }
	\label{Fig:PR_450GHz}
\end{figure}

\hspace{1em}We have also studied the performance of our detectors at frequencies intermediate to those reported in Figs. 2a and b of the main text and found that the resonant operation onsets already in the middle of the sub-THz domain. Figure \ref{Fig:PR_450GHz} shows the gate voltage dependence of $R\sous{a}$ recorded in response to 460 GHz radiation. In the vicinity of the charge neutrality point (NP) the responsivity peaks and changes its sign in agreement with the evolution of the FET-factor with the gate voltage (black curve in Supplementary Fig. \ref{Fig:PR_450GHz}) as discussed in the main text. However, away from the NP the responsivity peaks for both electron and hole doping (stars in Supplementary Fig. \ref{Fig:PR_450GHz}) despite the fact that $F(V\sous{g})$ is featureless. These peaks stem from the plasmon resonances in the FET channel as it follows from the comparison of the experimental data with theory (inset of Supplementary Fig. \ref{Fig:PR_450GHz}). In good agreement with theory, at lower frequencies the number of resonant modes, which can be observed for the same gate voltage span, is smaller compared to that found at 2 THz (Fig. 2b of the main text). In addition, the resonances appear much broader than those observed at 2 THz (Fig. 2b) which is consistent with the reduced quality factor at sub-THz frequencies. 

\subsection*{Supplementary Note 5: Fabri-Perot cavity model for plasmonic field-effect transistor} \label{Section: Fabri-Perot model}
\hspace{1em}Gated two-dimensional electronic systems support plasma waves with the dispersion relation  \cite{Chaplik1972}
\begin{equation}
	\label{Plasmon-dispersion}
	\omega (\omega + i \tau^{-1}) = \frac{n_0 e^2 q}{2 m^* \varepsilon_0 \varepsilon} (1 - e^{-2qd}).
\end{equation}
where $\omega$ and $q$ are the plasmon wavelength and wave vector, respectively, $\tau$ is the momentum relaxation time, $n_0$ is the carrier density, $m^*$ is the effective mass of charge carriers, $d$ it the distance to the gate, $\varepsilon$ is the dielectric permittivity, and $\varepsilon_0$ is the vacuum permittivity.

Confinement of a 2d channel by source and drain contacts quantizes the wave vector $q$ and leads to emergence of discrete plasmon frequencies. The quantization conditions can be obtained by requiring the oscillating quantity (e.g. voltage $V_\omega$) to return to its original value after the channel round trip:
\begin{equation}
	V_\omega r_s r_d   e^{2iqL} = V_\omega ,
\end{equation}
where $r_s$ and $r_d$ are the complex-valued reflection coefficients at the source and drain terminals, respectively. Therefore, eigen frequencies of bounded plasmons can be found from
\begin{equation}
	1 - r_s r_d   e^{2iqL} = 0.
\end{equation}
To see that the latter dispersion relation indeed appears in the nonlinear response functions, we model the FET channel as a transmission line (TL) fed by antenna voltage  $U_{1\omega}=V_a \cos\omega t$ at the source side~\cite{Aizin2012,dyer2013}. The antenna may have finite impedance $Z_{\rm a}$ which will be taken into account at the end of this section. The TL is terminated by load impedance $Z_{\rm gd}$ at the drain side, and by impedance $Z_{\rm gs}$ at the source side. In real device, these impedances are due to the capacitive coupling between the respective electrodes. The TL model is justified by the formal coincidence of TL equations (Telegrapher's equation) with transport equations in a gated FET channel.
\begin{figure}[ht!]
	\centering\includegraphics[width=0.5\linewidth]{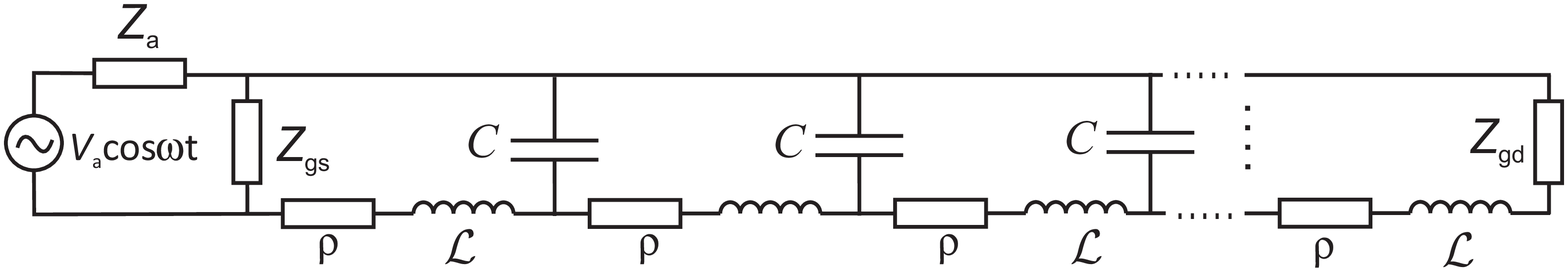}
	\caption{\textbf{Transmission line equivalent circuit of gated 2d channel.} $\mathcal L$ is the kinetic inductance of electrons, $\rho$ is the channel resistivity, $C$ is the effective gate-to-channel capacitance, and $Z_{\rm gd}$ is the load resistance at the drain side. All quantities are measured per unit length of the channel.  $Z_{\rm gd} \rightarrow \infty$ corresponds to Dyakonov-Shur boundary condition}
	\label{Supp-Equiv}
\end{figure}

The TL elements are specific inductance
\begin{equation}
	\label{Inductance}
	{\mathcal L} = \frac{m^*}{n_0 e^2 W}, 
\end{equation} 
capacitance per unit length
\begin{equation}
	\label{Capacitance}
	{C} =\frac{2 W \varepsilon \varepsilon_0 q}{1-e^{-2qd}},
\end{equation} 
and resistance
\begin{equation}
	\label{Resistance}
	\rho = {\mathcal L}/\tau,
\end{equation}
where $W$ is the channel width.
It is readily seen that the dispersion relation for waves in an infinite transmission line~\cite{collin1960} 
\begin{equation}
	q = \sqrt{C\mathcal{L}}\sqrt{\omega(\omega + \frac{i\rho}{\mathcal L})}
\end{equation}
coincides with plasma wave dispersion (\ref{Plasmon-dispersion}) with proper values of line parameters (\ref{Inductance}-\ref{Resistance}). The characteristic (wave) impedance of transmission line is
\begin{equation}
	Z_{\rm tl} = \sqrt{\frac{\omega\mathcal{L} + i \rho}{\omega C}}.
\end{equation}
A well-known result for current reflection coefficient from a loaded (drain) end of transmission line reads
\begin{equation}
	\label{DrainReflection}
	r_d = \frac{Z_{\rm tl} - Z_{\rm gd}}{Z_{\rm tl} + Z_{\rm gd}},
\end{equation}
while for source end with fixed voltage 
\begin{equation}
	\label{SourceReflection}
	r_s = 1.
\end{equation}

When the reflection coefficients and conditions at the ends of cavity are specified, it is straightforward to write down the solution for voltage across the TL (which is the gate-to-channel voltage in the actual FET):
\begin{equation}
	\label{Voltage-result}
	V_{\omega}(x) = \frac{V_a}{2} \frac{e^{iqx} - r_s r_d e^{-iq(x-2L)}}{1 - r_s r_d e^{2iqL}},
\end{equation}
here $r_d$ and $r_s$ are given by Eqs.~(\ref{DrainReflection}) and (\ref{SourceReflection}), respectively. The longitudinal electric field in the channel is given by
\begin{equation}
	\label{Field-result}
	E_{x\omega} = \frac{q V_a}{2} \frac{e^{iqx} +r_s r_d e^{-iq(x-2L)}}{1 - r_s r_d e^{2iqL}}.
\end{equation}
As the nonlinear response of the FET is proportional to the properly averaged square of ac electric field in the channel (\ref{Field-result}), it becomes apparent that responsivity would possess a plasma resonant factor $|1 - r_s r_d e^{2iqL}|^{-2}$, independent of the detection mechanism.

\addDS{The account of finite antenna impedance results in a a simple ''renormalization'' of input voltage in Eqs.~(\ref{Voltage-result}) and (\ref{Field-result}):
	\begin{equation}
		\label{Voltage_renormalized}
		V_a \rightarrow \frac{V_a}{1 + \frac{Z_a}{Z_{\rm gs}\parallel Z_{\rm tl, in}}},
	\end{equation}
	where $\parallel$ stands for parallel connection of impedances, and $Z_{\rm tl, in} = Z_{\rm tl}(1 - r_d e^{2iqL})/(1 + r_d e^{2iqL})$ is the input impedance of the transmission line (we have used $r_s = 1$). It is straightforward to show that the modification of input voltage can be translated in the modification of ''resonant denominator''
	\begin{equation}
		\label{Resonant_denominator}
		1-r_d e^{2iqL} \rightarrow (1-r_d e^{2iqL})\left[1+\frac{Z_a}{Z_{\rm gs}}\right]+(1+r_d e^{2iqL})\frac{Z_a}{Z_{\rm tl}}.
	\end{equation} 
	The effect of $Z_a$ in the square bracket is the reduction of input voltage due to the drop at internal antenna resistance. Finite value of $Z_a$ in the second term leads to extra broadening or resonances, as analyzed below. 
}  

\subsection*{Supplementary Note 6: Gate tuning of graphene plasmons: monolayer vs bilayer} \label{Section: Gate tuning}
We briefly review the density dependences of plasmon frequencies in single layer graphene (SLG) and bilayer  graphene (BLG). The general dispersion relation for gated plasmons in two-dimensional electron system with sheet conductivity $\sigma$ reads~\cite{Chaplik1972}
\begin{equation}
	\label{Plasmons_Conductivity}
	1 + \frac{i q \sigma}{2 \omega \epsilon \epsilon_0}(1-e^{-2qd})=0.
\end{equation}
The study of plasmon dispersions in various two-dimensional systems is therefore reduced to evaluation of their frequency-dependent conductivity. In the classical ($\hbar\omega\ll \varepsilon_F$) long-wavelength ($q\ll \omega/v_F$) limit, the latter is found from the Boltzmann equation~\cite{DasSarma2011}
\begin{equation}
	\sigma = \frac{e^2}{2}\int\limits_{0}^{+\infty }{d\varepsilon \rho(\varepsilon)\frac{v_p^2}{-i\omega +\tau _p^{-1}}\left( -\frac{\partial {{f}_{0}}}{\partial \varepsilon } \right)},
\end{equation}
where $\rho(\varepsilon)$ is the density of states, $v_p$ is the velocity of carrier with momentum $p$, and $f_0$ is the equilibrium distribution function. In case of BLG, 
$\rho(\varepsilon) = 2m/\pi \hbar^2$, $v_p = p/m$, which results in ordinary Drude conductivity 
\begin{equation}
	\label{SigmaBLG}
	\sigma_{\rm BLG} = \frac{ne^2\tau_p/m}{1-i\omega\tau_p}.
\end{equation}
In case of SLG, $\rho(\varepsilon) = 2\varepsilon/\pi\hbar^2v_F^2$, $v_p = v_F$, and the conductivity reads 
\begin{equation}
	\label{SigmaSLG}
	\sigma_{\rm SLG} = \frac{e^2}{\pi\hbar^2}\frac{kT\ln(1+e^{\varepsilon_F/kT})}{-i\omega+\tau_p^{-1}} \approx  \frac{e^2}{\pi\hbar^2}\frac{\varepsilon_F}{-i\omega+\tau_p^{-1}}.
\end{equation}
The latter equality is valid at low temperatures. Using the low-temperature relation between density and Fermi energy in SLG $n = \varepsilon_F^2/p\hbar^2v_F^2$, we readily observe that classical conductivity of SLG is still given by the Drude formula (\ref{SigmaBLG}) with density-dependent mass $m \rightarrow \varepsilon_F/v_F^2 \propto n^{1/2}$. Combining Eqs.~(\ref{Plasmons_Conductivity}), (\ref{SigmaBLG}) and \ref{SigmaSLG}, we observe that plasma frequency in BLG scales as $\omega\propto n^{1/2}$, while in SLG $\omega\propto n^{1/4}$.

\subsection*{Supplementary Note 7: Resonance broadening and plasmon lifetime} \label{Section: Broadening}

\hspace{1em}Before discussing the physics beyond THz rectification, we specify mechanism-independent quantities, namely, the positions of plasma resonances and resonance width. Introducing the complex reflection phase
\begin{equation}
	\exp\left[i \theta_r' - \theta''_r\right] = - r_s r_d,
\end{equation}
we transform the ''resonant denominator'' in eq. (\ref{Resonant-responsivity}) of the main text
\begin{equation}
	\label{Lineshape}
	R(\omega, V_g) \propto  |1 - r_s r_d e^{2iqL}|^{-2} = \frac{1}{2}\frac{e^{\theta''_r + 2q''L}}{\cosh (\theta''_r+2q''L) + \cos (\theta'_r+2q'L)},
\end{equation}
The maxima of responsivity correspond to wave vectors
\begin{equation}
	q'_{0} = \frac{\pi}{2L}(2k + 1 + \theta'_r/\pi).
\end{equation}
In the case of Dyakonov-Shur boundary conditions realized in our devices, $\theta_r = 0$, and the first resonance corresponds to $L$ equal to the quarter of plasmon wavelength. Assuming reflection and scattering losses to be small, the lineshape (\ref{Lineshape}) can be transformed to Lorentzian in the vicinity of each peak
\begin{equation}
	\label{Lineshape-resonant}
	R(\omega, V_g) \propto   \frac{e^{\theta''_r + 2q''L}}{(q''L +\theta''_r/2)^{2} + (q' - q_0)^2L^2}.
\end{equation}
The full width at half-height is given by
\begin{equation}
	\label{Width-wavevector}
	\frac{\delta q}{q_0} = \frac{1}{\omega \tau} +\frac{2}{\pi}\frac{\ln |r_s r_d|^{-1}}{2k+1} \equiv \frac{1}{\omega \tau_{p}},
\end{equation}
here we have introduced the plasmon lifetime $\tau_{p}$ which is below the scattering time $\tau_p$ due to resonator loss. \addDS{It is also possible  to take into account the effect of finite antenna resistance on plasmon linewidth. To this end, one should transform resonant denominator of the form (\ref{Resonant_denominator}) in the vicinity of resonance. This leads us to
	\begin{equation}
		\frac{1}{\omega\tau_p} = \frac{1}{\omega \tau} + \frac{2}{\pi}\frac{\ln |r_s r_d|^{-1}+2 Z'_{\rm a}/Z'_{\rm tl}}{2k+1}.
	\end{equation}
	The above equation clearly demonstrates that inverse plasmon lifetime $\tau^{-1}_{\rm p}$ is the sum of electron momentum relaxation rate $\tau^{-1}$, contact damping rate \begin{equation}
		\tau^{-1}_{\rm cont} = \frac{2\omega}{\pi}\frac{\ln |r_s r_d|^{-1}}{2k+1} = \frac{s}{L}\ln|r_sr_d|^{-1}\propto \sqrt{V_g},
	\end{equation}
	and damping rate due to antenna resistance
	\begin{equation}
		\tau^{-1}_{\rm ant} = \frac{2s}{L}\frac{Z'_{\rm a}}{Z'_{\rm tl}}\propto V_g.
	\end{equation}
	The two latter contributions to damping rate are minimized in the vicinity of charge neutrality point. Examples of calculated plasmon lifetime including the contributions of contacts and antenna are shown in Supplementary Fig.~\ref{Plasmon_lifetime}. Effects of  radiative contribution to plasmon damping on detector responsivity is shown in Supplementary Note 9, along with the discussion of detection mechanisms.}

\begin{figure}[ht!]
	\centering\includegraphics[width=0.4\linewidth]{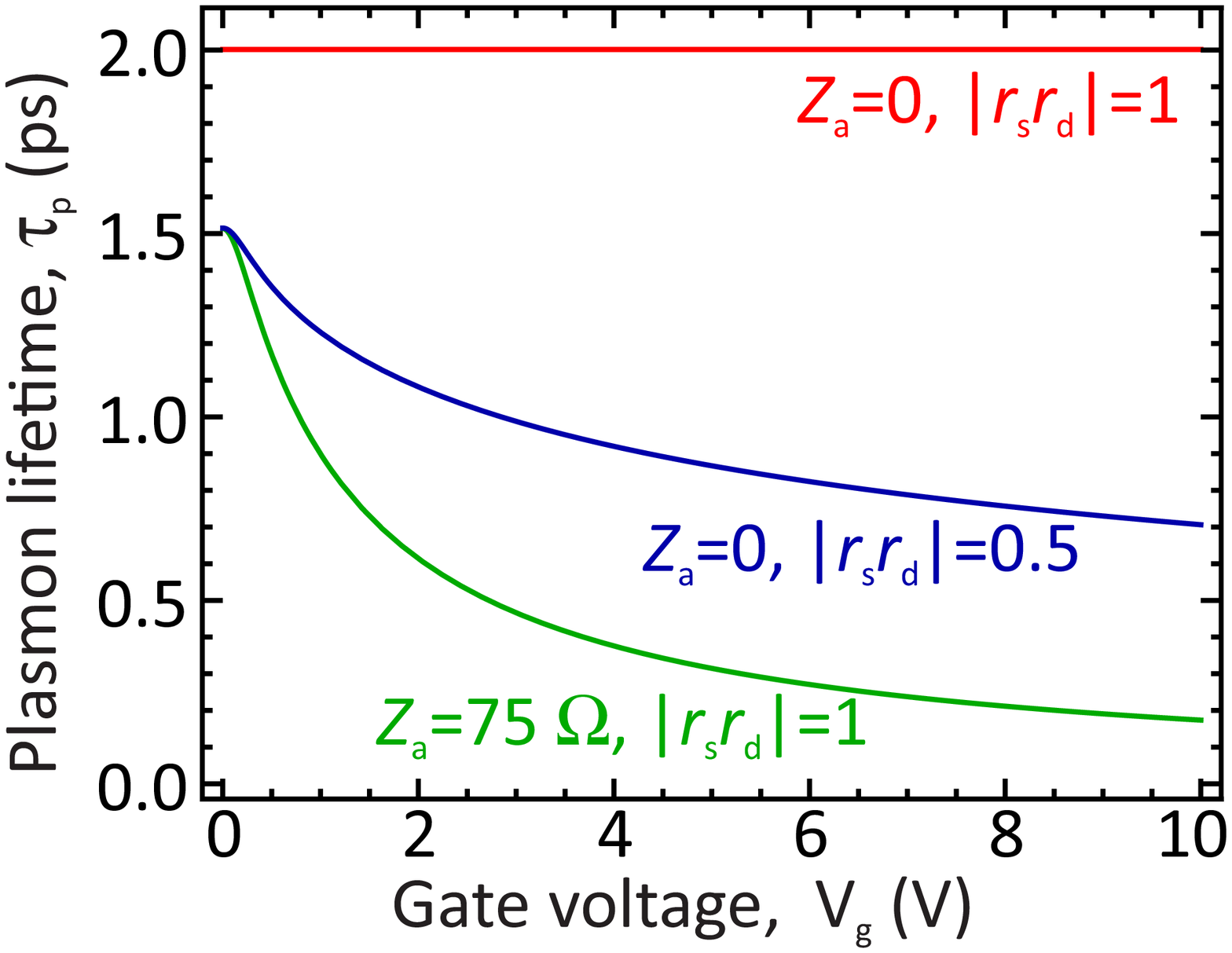}
	\caption{\textbf{Plasmon damping.} Calculated plasmon lifetimes assuming momentum relaxation time $\tau=2$ ps and no antenna and contact losses (red), lossless antenna and reflection coefficient $|r_sr_d| = 0.5$ (blue), antenna resistance $Z'_a = 75$ Ohm and perfect reflection (green). Plasmon velocity $s$ at charge neutrality point is limited by residual carriers with density $n^* =5\times 10^{10}$ cm$^{-2}$.}
	\label{Plasmon_lifetime}
\end{figure}

As the wave vector at fixed frequency is inversely proportional to wave velocity, $q =\omega/s\propto V_g^{-1/2}$, expression (\ref{Width-wavevector}) can be transformed to the voltage scale
\begin{equation}
	\label{Width-voltage}
	\frac{\delta V_g^{-1/2}}{V_g^{-1/2}} = \frac{1}{\omega \tau_{p}}.
\end{equation}

\subsection*{Supplementary Note 8: Photothermoelectric rectification in Fabri-Perot cavity} \label{Section: PTE}
\hspace{1em}	Asymmteric feeding of THz radiation results in asymmetric heating of the device and emergence of thermoelectric effect. The resulting dc voltage is~\cite{Bandurin2017}
\begin{equation}
	eV_{\rm pte} = (S_{ch} - S_{cont})\left[ T_s - T_d \right],
\end{equation}
where $S_{ch}$ is the Seebeck coefficient in the gated channel, and $S_{cont}$ -- in the metal-doped graphene contact, $T_s$ is the local temperature at the source junction and $T_d$ is at the drain junction 
. From now on, we refer to the gated part of graphene as ''channel'' and ungated part -- as ''contact''. The doping of ungated part does not depend on gate voltage, however, it can be non-uniform due to the effects of built-in field near metal contacts.

The temperature difference $T_s-T_d$ induced by non-uniform heating of the device can be found from the solution of heat transfer equation in the channel:
\begin{gather}
	\label{Heat-transfer}
	\frac{\partial^2 T}{\partial x^2} + \frac{T - T_0}{L^2_T} = -\frac{q(x)}{\chi_{ch}},\\
\end{gather}
where $q(x) = 2 {\rm Re}\sigma_\omega |E_{x\omega}|^2$ is the Joule heating power, $\chi_{ch}$ is the electron thermal conductivity in the channel, $L_T = (\chi_{ch} \tau_\varepsilon/C_e)^{1/2}$ is the thermal relaxation length, $\tau_\varepsilon$ is the energy relaxation time due to heat sink into substrate phonons~\cite{Low2012}, and $C_e$ is the heat capacitance of the electronic system. Equation (\ref{Heat-transfer}) is supplemented by the boundary conditions at the boundaries of gated domain
\begin{equation}
	\frac{\chi_{cont}}{L_{cont}}(T_s - T_0) = \chi_{ch} \nabla T_s,\qquad \frac{\chi_{cont}}{L_{cont}}(T_d - T_0) = - \chi_{ch} \nabla T_d;
\end{equation}
these conditions follow from the continuity of heat flux at the interfaces. The sought-for temperature difference between source and drain can be obtained in the closed form under the following simplifying assumptions (1) the Joule heating occurs only in the channel (2) the temperature drop across the contacts is much less than maximum overheating in the channel. Both conditions are justified by the small length of the contacts $L_{cont}\ll L$. Under these assumptions, the expression for the photo-thermoelectric voltage acquires a physically appealing form
\begin{equation}
	\label{PTE-final}
	eV_{\rm pte} = (S_{ch} - S_{cont})  \frac{l_{cont}}{\chi_{cont}}\int^L_0{
		2{\rm Re}\sigma_{\omega} |E_{x\omega}|^2 \frac{\sinh\frac{x-L/2}{L_T}}{\sinh \frac{ L}{2L_T}} dx}.
\end{equation}

The quantity $\chi_{cont}/l_{cont}$ is the thermal {\it conductance} of the contact, while the integral is the difference of heat fluxes traveling toward the source and toward the drain. The kernel of the integral is anti-symmetric with respect to the middle of the channel $x=L/2$, therefore, the PTE signal appears only due to asymmetric heating $q(x)$. Final evaluation of PTE voltage is performed by substituting the solution for electric field (\ref{Field-result}) into (\ref{PTE-final}):
\begin{multline}
	\label{PTE-Final}
	eV_{\rm pte} = (S_{ch} - S_{cont}) \frac{2L_{cont} L_T {\rm Re}\sigma_{\omega} }{\chi_{cont}}\frac{|q|^2V_a^2}{e^{q''L+\theta''_r/2} |1-r_s r_d e^{2iqL}|^2}\times\\
	\left[\sin2\alpha' \frac{2q'L_T \cos q'L - \coth\frac{L}{2L_T} \sin q'L }{1 + (2q' L_T)^2} + \sinh 2\alpha'' \frac{2q''L_T \cosh q''L - \coth\frac{L}{2L_T} \sinh q''L }{1 - (2q'' L_T)^2}\right],
\end{multline}
where $\alpha = \theta_r + qL/2$.

\subsection*{Supplementary Note 9: Dyakonov-Shur rectification in Fabri-Perot cavity} \label{Section: DS}
\hspace{1em} The so-called Dyakonov-Shur rectification includes two physically different nonlinearities. One contribution to the rectified current appears due to simultaneous modulation of 2d channel conductivity and application of longitudinal field. This effect, also known as resistive self-mixing, results in the rectified voltage
\begin{equation}
	\label{Self-mixing}
	V_{\rm rsm} = 2{\rm Re} \int_{0}^{L}{\frac{1}{\sigma_{\omega=0}}\frac{d \sigma_{\omega}}{dV_{\rm gc}}  V_{-\omega}(x) \frac{\partial V_{\omega}(x)}{\partial x}dx}.
\end{equation}
Here $\sigma_{\omega=0} = ne^2\tau/m$ is the dc conductivity of a 2D channel, $n$ is the carrier density, $e$ and $m$ are the elementary charge and effective mass of charge carriers respectively, $\tau$ is the momentum relaxation time, and $\sigma_\omega = \sigma_{\omega=0}/(1-i\omega \tau)$ is the high-frequency conductivity. Evaluation of the integral leads us to the result
\begin{equation}
	\label{RSM-final}
	V_{\rm rsm} =
	\frac{V_{a}^2/V_{g}}{\sqrt{1+\omega^2\tau^2}}\frac{\frac{q''}{q'} [\cos\theta'_r - \cos(\theta'_r+2q'L)] +\frac{q'}{q''} [\cosh\theta''_r - \cosh(\theta''_r+2k''L)]}{2e^{\theta''_r+2q''L}|1-r_s r_d e^{2iqL}|^2}
\end{equation}

\begin{figure}[ht!]
	\centering\includegraphics[width=0.4\linewidth]{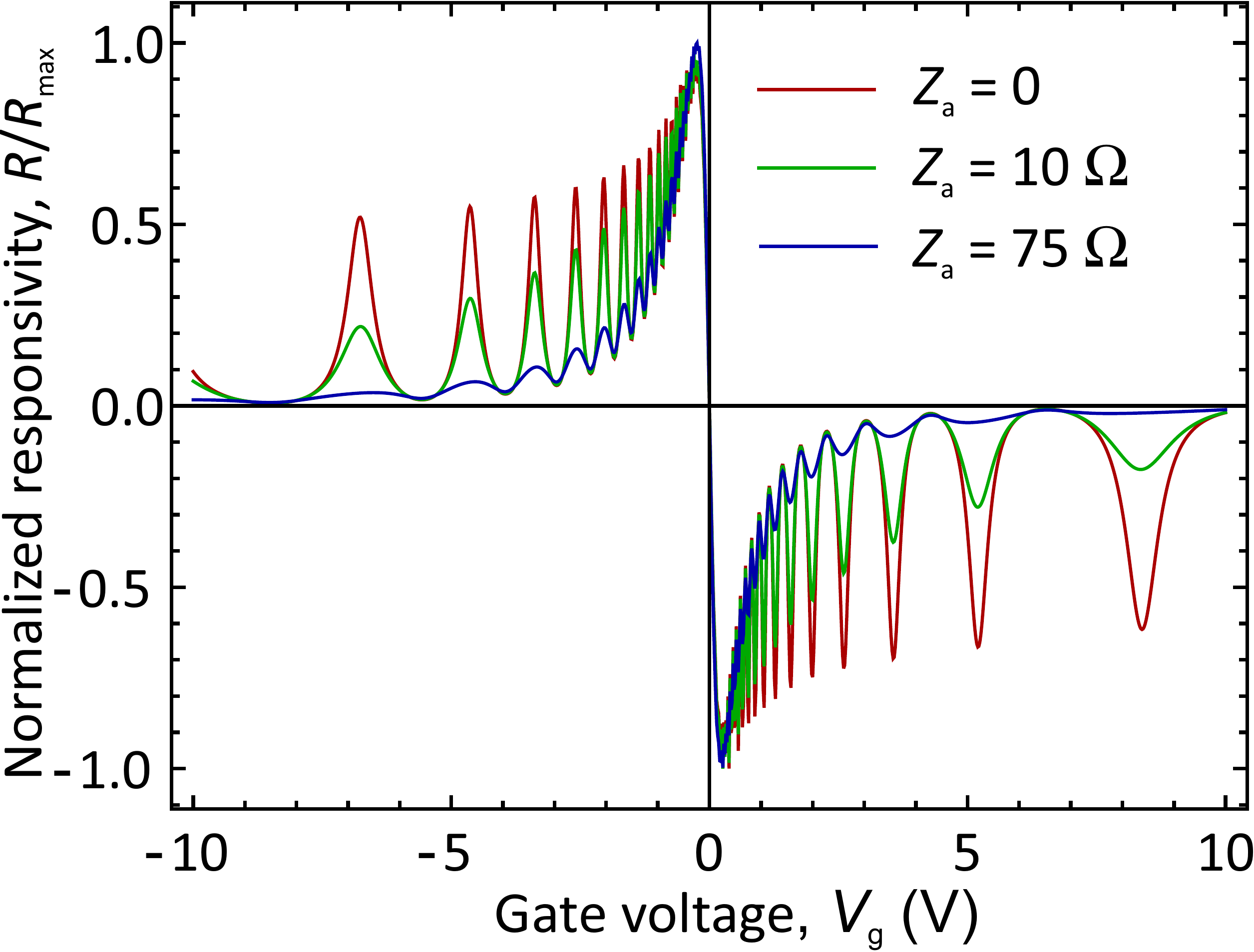}
	\caption{\textbf{Dyakonov-Shur photoresponse.} \addDS{Calculated Dyakonov-Shur responsivity of BLG FET detector as a function of gate voltage at different values of antenna impedance. All curves are normalized by their maximum value reached in the vicinity of NP. Parameters: channel length $L=6$ $\mu$m, momentum relaxation time $\tau=2$ ps, radiation frequency $f=2$ THz, gate-channel separation $d=80$ nm, residual carrier density $n^* = 5\times 10^{10}$ cm$^{-2}$.}}
	\label{Antenna_resonances}
\end{figure}

Another contribution to rectified voltage stems from the difference of kinetic energies of electron fluid at the source and drain side (Bernoulli law). The underlying nonlinearity is manifested by convective term $({\bf u \nabla}) {\bf u}$ in the Euler equation for electron fluid~\cite{Dyakonov1996a}. The corresponding rectified voltage is 
\begin{equation}
	\label{Nonlinear}
	e V_{\rm nl} = \frac{e^2}{m^*(\omega^2 + \tau^{-2})}\left[|E_{x\omega}(L)|^2 - |E_{x\omega}(0)|^2 \right],
\end{equation}
where $E_{x\omega}$ is the complex amplitude of high-frequency longitudinal field in the channel given by (\ref{Field-result}). Using the result for electric field (\ref{Field-result}), we find
\begin{gather}
	\label{Nonlinear-final}
	V_{\rm nl} = \frac{1}{4} \frac{V_{a}^2/V_{g}}{\sqrt{1+\omega^{-2}\tau^{-2}}}
	\frac{\left|1+r_s r_d e^{2iqL}\right|^2 - \left|(1+r_s r_d)e^{iqL}\right|^2 }
	{\left|{1-r_s r_d e^{2iqL}}\right|^2}.
\end{gather}
Equations (\ref{RSM-final}) and (\ref{Nonlinear-final}) generalize the known results of Dyakonov and Shur for arbitrary loading of the plasmonic FET at the terminals. Naturally, they reduce to the result of Ref.~\citen{Dyakonov1996a} for high-impedance drain load $\theta_r=0$, yielding the photovoltage given by:
\begin{equation}
	\label{DS_from_the_paper}
	\Delta U= -\frac{1}{4} \frac{V_a^2}{V_g} \left[1 + \frac{2}{\sqrt{1+(\omega \tau)^{-2}}} - \frac{1 + \frac{2 \cos 2q'L}{\sqrt{1+(\omega\tau)^{-2}}}}{\sinh^2q''L + \cos^2 q'L}\right],
\end{equation}
The factor in square brackets peaks when the length of the FET channel matches odd multiples of the plasmon quarter-wavelength.

We note that Eqs. (\ref{RSM-final}), (\ref{Nonlinear-final}) and (\ref{DS_from_the_paper}) diverge as the dc gate voltage $V_g$ tends to zero. In fact, this divergence stems from the gradual-channel approximation, relating carrier density and gate voltage $CV_g = e n$, that fails near the charge neutrality point. The account of ambipolar transport involving electrons and holes leads to a simple replacement in Eqs. (\ref{RSM-final}), (\ref{Nonlinear-final}) and (\ref{DS_from_the_paper}):
\begin{equation}
	\label{Eq:El-holes}
	\frac{1}{V_g} \rightarrow \frac{n/m_n^2 - p/m_p^2}{n/m_n + p/m_p}\frac{1}{s^2},
\end{equation}
where $n$ and $p$ are electron and hole densities, $m_n$ and $m_p$ are their effective masses, and $s^2 = (n/m_n+p/m_p)e^2/C$ is the plasma wave velocity in ambipolar system. 

In the main text, the experimental photoresponse was compared with the DS photovoltage from eq. (\ref{DS_from_the_paper}) corrected by eq. (\ref{Eq:El-holes}). \addDS{The full picture of calculated Dyakonov-Shur photovoltage vs gate voltage is shown in Supplementary Fig.~\ref{Antenna_resonances}. Along with the result for perfect reflection from the drain and ideal voltage source ($Z_{\rm a}=0$, red line), it also shows the effect of finite antenna impedance on resonance width (green and blue lines). In accordance with the discussed antenna-induced ''renormalization'' of input voltage, eq.~\ref{Voltage_renormalized}, the resonances at high carrier density are highly broadened due to finite $Z_{\rm a}$. The width of resonances at low density, on the contrary, is mainly determined by momentum relaxation time.}

\subsection*{Supplementary Note 10: Detector operation outside the linear regime} \label{Section: nonlinear}
\hspace{1em} The data reported in the main text were obtained in the regime where the detector's photovoltage grew linearly with the power $P$ of incoming radiation. With increasing $P$ outside the linear regime, the heating of graphene's electronic system by high-frequency ac capacitive currents flowing between the source and gate terminals may affect the resulting responsivity. In order to reveal the role of heating, we have studied the response of our detectors at varying $P$ and found that outside the linear regime, $R\sous{a}$ decreases with increasing $P$. We attribute this decrease to the modification of the channel conductivity with increasing electronic temperature. To support this statement, we plot the FET-factor obtained by measuring the sample’s conductivity at different $T$ inside the sample chamber. Clearly, $R\sous{a}$ acquired at different $P$ follows the evolution of the FET-factor $F$ with $T$. This is reflected in the shift and decrease of the responsivity extrema with increasing $P$. We also refer to Fig. 2a of the main text which shows $R\sous{a}(V\sous{g})$ at different $T$ that resembles the behaviour shown in Supplementary Fig. \ref{FIG_heating}.

\begin{figure}[ht!]
	\centering\includegraphics[width=0.4\linewidth]{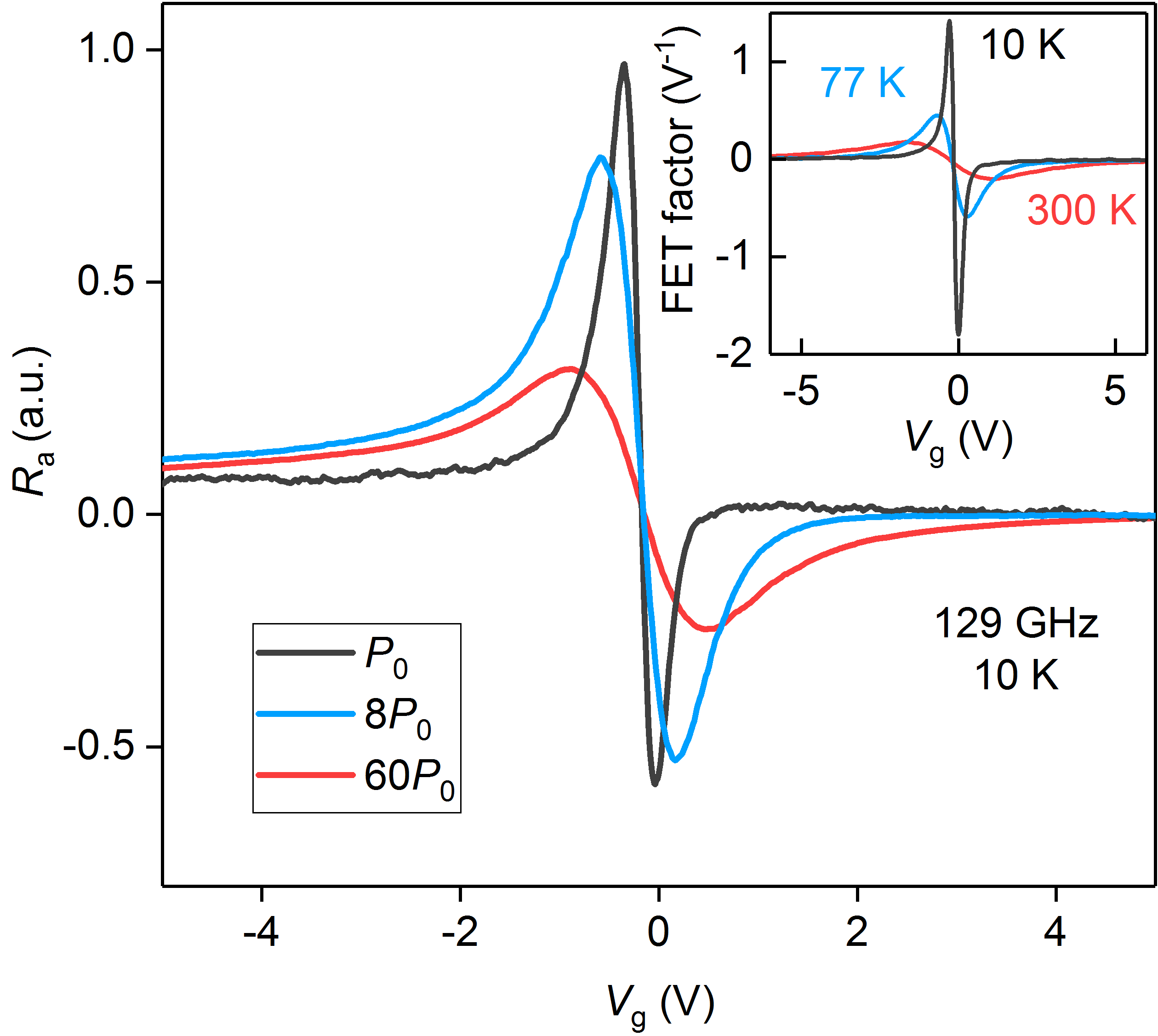}
	\caption{\textbf{Role of electron heating.} Normalized to unity responsivity as a function gate voltage measured in one of our BLG detectors at given $T$ and $f$ for different $P$. $P\sous{0}=4$ $\mu$W. Inset: FET-factor as a function of $V\sous{g}$ for different $T$.}
	\label{FIG_heating}
\end{figure}


\end{document}